\documentclass[11pt,letterpaper]{article}
\pdfoutput=1
\usepackage{jcappub}
\usepackage{color}
\usepackage{graphicx}
\usepackage{wrapfig}

\usepackage{cancel}

\usepackage{verbatim}
\usepackage{amsmath}
\usepackage{amssymb}
\usepackage{subfig}
\usepackage{url}
\usepackage{bbold}
\usepackage{xspace}

\usepackage{multirow}
\usepackage{threeparttable}
\usepackage{paralist}

\usepackage{color}
\definecolor{darkgreen}{rgb}{0,0.5,0}

\newcommand{\U}{\text{U}}
\newcommand{\SU}{\text{SU}}

\DeclareRobustCommand{\Sec}[1]{Sec.~\ref{#1}}

\DeclareRobustCommand{\App}[1]{App.~\ref{#1}}
\DeclareRobustCommand{\Tab}[1]{Table~\ref{#1}}

\DeclareRobustCommand{\Fig}[1]{Fig.~\ref{#1}}

\DeclareRobustCommand{\Eq}[1]{Eq.~(\ref{#1})}

\DeclareRobustCommand{\Refs}[1]{Refs.~\cite{#1}}

\newcommand{\be}{\begin{equation}}
\newcommand{\ee}{\end{equation}}

\newcommand{\mb}[1]{\boldsymbol{#1}}

\begin{document}

\title{Exothermic Double-Disk Dark Matter}

\author[a]{Matthew McCullough}
\author[b]{and Lisa Randall}

\affiliation[a]{Center for Theoretical Physics, Massachusetts Institute of Technology,\\Cambridge, MA 02139, USA}
\affiliation[b]{Department of Physics, Harvard University, Cambridge, MA 02138, USA}

\emailAdd{mccull@mit.edu}
\emailAdd{randall@physics.harvard.edu}

\date{\today}


\arxivnumber{}

\abstract{If a subdominant component of dark matter (DM) interacts via long-range dark force carriers it may cool and collapse to form complex structures within the Milky Way galaxy, such as a rotating dark disk.  This scenario was proposed recently and termed ``Double-Disk Dark Matter'' (DDDM).  In this paper we consider the possibility that DDDM remains in a cosmologically long-lived excited state and can scatter exothermically on nuclei (ExoDDDM).  We investigate the current status of ExoDDDM direct detection and find that ExoDDDM can readily explain the recently announced $\sim 3 \sigma$ excess observed at CDMS-Si, with almost all of the $90\%$ best-fit parameter space in complete consistency with limits from other experiments, including XENON10 and XENON100.  In the absence of isospin-dependent couplings, this consistency requires light DM with mass typically in the $5-15$ GeV range.  The hypothesis of ExoDDDM can be tested in direct detection experiments through its peaked recoil spectra, reduced annual modulation amplitude, and, in some cases, its novel time-dependence.  We also discuss future direct detection prospects and additional indirect constraints from colliders and solar capture of ExoDDDM.   As theoretical proof-of-principle, we combine the features of exothermic DM models and DDDM models to construct a complete model of ExoDDDM, exhibiting all the required properties.}

%\preprint{MIT-CTP {4480}}

\maketitle

\section{Introduction}
\label{sec:introduction}
The case for physics beyond the Standard Model, in the form of non-baryonic Dark Matter (DM), is overwhelming.  The evidence, based solely on gravitational interactions, has cast light on the large-scale properties of the dominant DM components.  However the detailed composition and dynamics of DM remains a mystery and a particle physics interpretation would yield a true understanding of the majority of matter in the Universe.  Exploration of the particle nature of DM has typically progressed through the complementary theoretical and experimental study of certain well-motivated DM candidates, leading to specific experiments targeted at detecting the interactions expected.  However, as these DM candidates and the overarching frameworks within which they reside remain undiscovered it is becoming increasingly important to broaden theoretical horizons and to cast a wider experimental net.

One often under-appreciated possibility for DM behavior is that, although we know the majority of DM should be currently cold and collisionless, there may exist rich and varied subcomponents of the DM budget with a host of unexpected properties.  In fact, we already know that a small component of the total matter budget exhibits such behavior: the baryonic matter that makes up the visible Universe.  This possibility, which was recently emphasized and explored in \Refs{Fan:2013yva,Fan:2013tia}, could have profound consequences for the detection of DM and is deserving of serious consideration.

The possibilities for complex dynamics in a subcomponent of DM are vast.  As well as pointing out that DM can have this richer structure, \Refs{Fan:2013yva,Fan:2013tia} highlighted one particularly interesting scenario.  If some subcomponent of the DM has significant self-interactions and can cool sufficiently rapidly through scattering and emission of very light, or massless, dark states then this subcomponent could collapse to form structures similar to those observed in the visible sector.\footnote{See also \cite{darkint} for early work concerning long-range dark forces.}  In particular this may lead to the formation of galactic dark disks, coexisting within visible galactic baryonic disks.  This scenario was termed ``Double-Disk Dark Matter'' or ``DDDM'' for short \cite{Fan:2013yva,Fan:2013tia}.  It should be emphasized that this extra subcomponent of DM is distinct from the dominant cold and collisionless component which makes up the majority of the DM.\footnote{It should be noted that the dominant component of DM may also form a disk-like structure \cite{Read:2008fh,Read:2009iv,Purcell:2009yp,Bruch:2009rp} which may influence direct detection signals \cite{MarchRussell:2008dy,Ling:2009eh,Ling:2009cn,Green:2010gw,Billard:2012qu}.  However, in the DDDM scenario the dark disk is an entirely different species of particle from that which comprises the main DM halo.}

The DDDM may have interesting non-gravitational interactions with visible-sector particles and can lead to enhanced indirect detection signals, which could be distinguished from conventional DM scenarios as the spatial distribution of the indirect signal could be significantly different for DDDM \cite{Fan:2013yva,Fan:2013tia}.  However, as described in \cite{Fan:2013yva,Fan:2013tia}, the direct detection prospects are limited for elastically scattering DDDM.  This is because the disk of DDDM is expected to rotate with a comparable velocity to the visible baryonic disk.  Hence the average relative velocity between DM and an Earth based detector is small, leading to nuclear recoil energies below typical detector thresholds.  There is some relative velocity due to the peculiar velocity of the Sun, the orbit of the Earth, and the velocity dispersion of DDDM.  However these components combined are expected to be much smaller than the typical relative velocity expected for standard halo DM.

\begin{figure}[t!]
  \centering
  \includegraphics[width=0.45\textwidth]{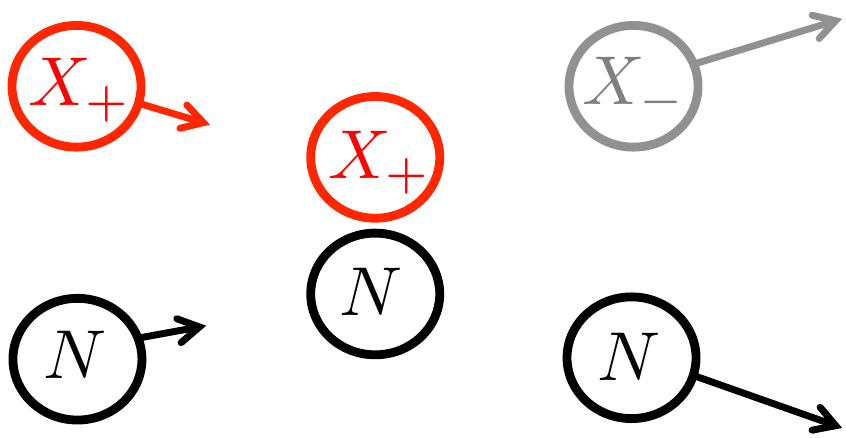}
    \caption{A schematic diagram of an exothermic DM-nucleus scattering event where speeds are indicated by the length of arrow.  An incoming DM particle is in an excited state and de-excites upon scattering at the centre.  The exothermic energy release depends on the mass-splitting between DM states, $M_+-M_- = \delta$, some of which is deposited as nuclear recoil energy.  Due to the exothermic nature of the scattering, nuclear recoils are possible even if the scattering event occurs at rest, or at low relative velocity, enabling DDDM direct detection; an otherwise challenging prospect.} \label{fig:exoDM}
\end{figure}

However, this does not rule out the possibility of DDDM direct detection signals.  In future experiments perhaps dedicated low-threshold analysis could detect scattering events.  Alternatively, if the DDDM is very massive then the kinetic energy is increased and could potentially lead to observable signatures above the low energy thresholds \cite{HeavyDDDM}.

In this paper we consider a modified scenario that can give rise to direct detection signals in current and planned experiments even in a DDDM context.  This scenario involves exothermic DDDM scattering on nuclei \cite{Graham:2010ca}.  In exothermic DM-nucleus scattering an excited DM state collides with a nucleus at which point the DM de-excites, depositing some part of the DM kinetic energy plus an additional component proportional to the mass-splitting of DM states, $M_+-M_- = \delta$.  The energy deposit is manifest in the detector as nuclear recoil energy.  This process is depicted in \Fig{fig:exoDM}.  Although the final result (a recoiling nucleus) is similar to the result of standard elastic DM-nucleus scattering, the kinematics are quite distinct and the typical energy deposited does not depend on the phase space distribution of DM in the usual way, leading to a novel recoil spectrum.

To enable clarity in comparisons between the phenomenology of DDDM and standard cold and collisionless DM we will refer to the exothermic scattering of DDDM as ExoDDDM and the exothermic scattering of a standard halo DM candidate as ExoDM.  In this work we will demonstrate that ExoDDDM may exhibit novel and distinctive energy spectra and modulation characteristics in direct detection experiments.\footnote{Throughout we will assume that the component of DM which is cold and collisionless, and dominates the energy density of DM does not lead to detectable scattering events in direct detection experiments.}  In \Sec{sec:pheno} we will outline the main qualitative features which distinguish direct detection signals of ExoDDDM from ExoDM or elastic DM, and show that it may be possible to infer that DM signals are coming from a DDDM subcomponent rather than a standard DM candidate.  In \Sec{sec:quantitative} we will make these arguments quantitative by considering current direct detection limits on ExoDDDM.  We will also demonstrate that the three candidate events recently reported by the CDMS collaboration \cite{Agnese:2013rvf} can be readily explained by ExoDDDM scattering, in complete consistency with limits from other detectors.  In particular, for certain values of the exothermic splitting there is virtually no tension between the strongest limits from the XENON10 and XENON100 experiments and the majority of the $90\%$ best-fit parameter space for an ExoDDDM explanation of the CDMS-Si events, even when the coupling of ExoDDDM to protons and neutrons is equal.  We will also discuss the possibility of concurrently explaining the DAMA, CoGeNT and CRESST-II anomalies, finding that this is not possible for all four experiments with ExoDDDM.  However consistent interpretations of CDMS-Si and CRESST-II excesses may be possible.  Variations of the proton and neutron couplings \cite{Kurylov:2003ra,Giuliani:2005my,Chang:2010yk,Feng:2011vu,Feng:2013vod} are also considered in the context of ExoDDDM and new high-mass explanations of the CDMS-Si events are found where $M \lesssim 80$ GeV, in consistency with other bounds.  In \Sec{sec:colliderindirect} we consider collider and solar capture constraints on ExoDDDM.  In \Sec{sec:model} we construct a complete model of ExoDDDM which exhibits the properties required for cooling and collapse into DDDM as well as the mass splitting and long-lived excited state required for exothermic scattering on nuclei.  We conclude in \Sec{sec:conclusions}.

\section{Direct Detection Phenomenology}
\label{sec:pheno}
Before considering the current experimental status of ExoDDDM in \Sec{sec:quantitative} it is first useful to outline some distinguishing qualitative features of ExoDDDM direct detection.\footnote{See also \cite{Graham:2010ca} for discussions of ExoDM scattering.}  These features arise due to a combination of effects from the exothermic scattering and the DDDM phase space distribution.

We begin with the features specific to exothermic scattering, first focussing on comparing scattering rates at different detectors.  The phenomenology can be best understood by considering the minimum velocity an incoming DM particle must have, $v_{\text{min}}$, to generate a measurable energy deposit in the detector, $E_R$. If the DM down-scatters on a nucleus as $X_+ + N \rightarrow X_- + N$ where $M_{\pm} = M \pm \delta/2$, then the minimum velocity any incoming DM particle must have to produce a given nuclear recoil energy, $E_R$, is
\be
v_{\text{min}}  (E_R) = \frac{1}{\sqrt{2 M_N E_{R}}} \left| \frac{M_N E_{R}}{\mu_{N}} - \delta \right|  ~~,
\label{eq:vmin}
\ee
where $\mu_{N}$ is the reduced mass of the DM-nucleus system and we have assumed $|\delta| \ll M$.

The local velocity distribution of DM is assumed to be the same at each detector.  For a given nuclear recoil energy \Eq{eq:vmin} demonstrates that there is a detector-dependent DM velocity threshold below which scattering cannot lead to detectable scattering events.  Thus the sensitivity of a detector depends on the experimental low energy threshold since this determines the fraction of the total DM distribution which leads to detectable scattering events.  Let us consider an idealized situation where there are detectors with different nuclear targets, but with the same low-energy threshold, $E_{\text{thr}}$, and no high energy threshold.\footnote{This is not the situation in reality, but serves to delineate the comparison between experiments.  For the sake of simplicity we also ignore finite resolution effects in this general discussion.}  In this case a given detector is sensitive to all DM particles with velocity greater than
\[v_{\text{thr}} = \left\{
  \begin{array}{l l}
    v_{\text{min}} (E_{\text{thr}}) & \quad \delta < E_{\text{thr}} M_N/\mu_N \\
    0 & \quad \delta > E_{\text{thr}} M_N/\mu_N
  \end{array} \right.\]

The latter arises since if $\delta > E_{\text{thr}} M_N/\mu_N$ there exists some recoil energy within the detector range, for which $v_{\text{min}}  (E_R)=0$.  This means that the detector is sensitive to the full velocity distribution of the DM, even though $v_{\text{min}}  (E_{\text{thr}}) > 0$.  For a given detector only DM particles with velocity $v_{\text{thr}}$ or above will be capable of producing a measurable signal.  Let us consider the dependence of $v_{\text{thr}}$ on the nuclear mass for the major categories of models:
\begin{itemize}
\item  Elastic, heavy DM:  In this case $v_{\text{thr}} \approx \sqrt{E_{\text{thr}}/2 M_N}$.  Heavier nuclei lead to reduced minimum velocity thresholds and will thus sample more of the DM velocity distribution, leading to greater sensitivity.
\item  Elastic, light DM:  In this case $v_{\text{thr}} \approx \sqrt{E_{\text{thr}} M_N}/\sqrt{2}M$, the minimum velocity threshold is reduced for lighter nuclei, and detectors with lighter nuclei will sample more of the DM velocity distribution, improving the sensitivity to light DM.
\item  Exothermic:  To the minimum velocity of the previous two elastic scattering cases we subtract an additional component such that $v_{\text{thr}} = |v_{\text{thr}}({\delta=0})-\delta/\sqrt{2 M_N E_{\text{thr}}}|$.  This extra exothermic term leads to a reduction in the minimum velocity, and can in some cases reduce it to zero.  The reduction is greatest for light nuclei, leading to preferential scattering of the DM on lighter target nuclei.  For light exothermic DM this further enhances the sensitivity of light nuclei detectors over heavy nuclei detectors.
\end{itemize}
Thus we see that with exothermic DM scattering the minimum DM velocity threshold is reduced relative to an elastically scattering DM candidate.  Importantly, this opens up the possibility of detecting DDDM in \emph{current} direct detection experiments since in some cases scattering events can occur above thresholds even if the DDDM is essentially stationary relative to the DM detector.  For light ExoDDDM there is also a clear preference for lighter target nuclei, which allows an ExoDDDM interpretation of the three events recently reported by CDMS-Si ($A_{\text{Si}} = 28$), completely consistent with bounds from the XENON ($A_{\text{Xe}} = 131$) experiments.

  \begin{figure}[t!]
  \centering
  \includegraphics[height=0.44\textwidth]{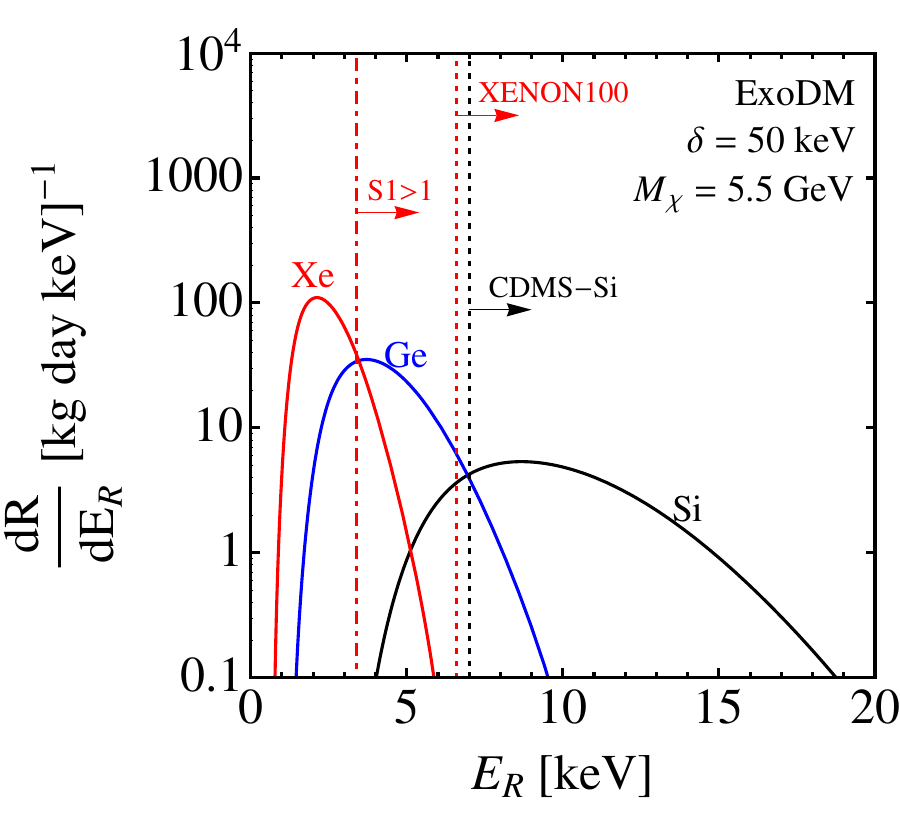}   \includegraphics[height=0.44\textwidth]{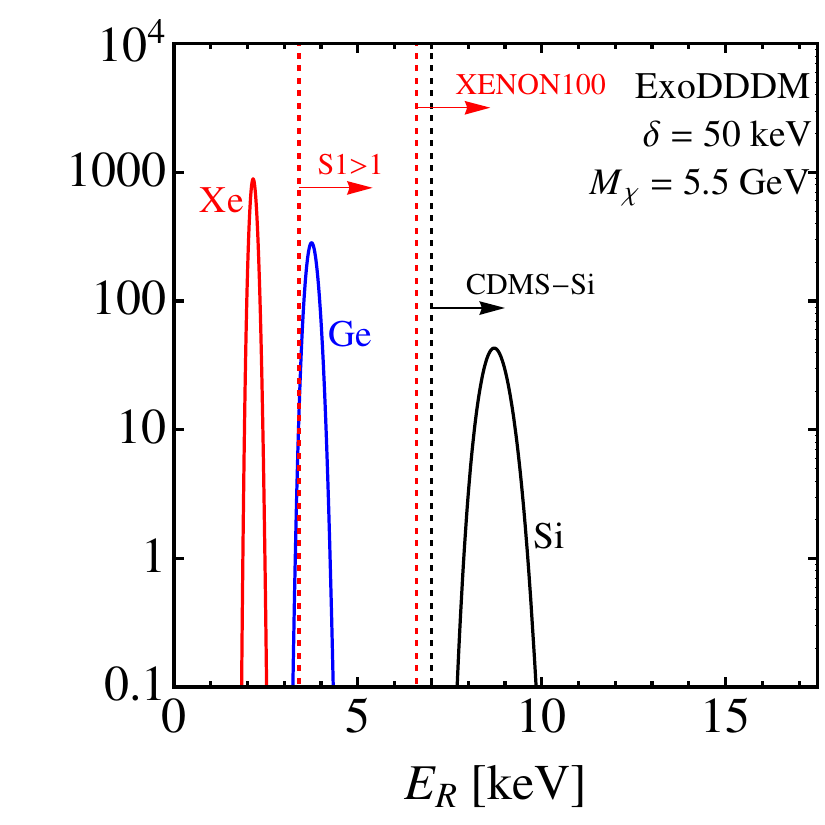}
  \caption{Low energy threshold for the CDMS-Si experiment (dotted black) and the low energy threshold and single photelectron (S1) threshold for XENON100 (dotted and dot-dashed red respectively).  Nuclear recoil energy spectra are also shown for ExoDM (left) and ExoDDDM (right) scattering on various nuclei for two benchmark parameter points.  As the mass of a nucleus is increased the typical recoil energies are driven to lower values, and can be below detector thresholds.  The recoil spectrum vanishes at low energies and exhibits a peak, in contrast to elastically scattering DM which shows an exponential rise towards lower energies.  A significant difference between ExoDM and ExoDDDM is that, due to the small velocity dispersion and relative velocity, ExoDDDM scattering leads to very narrow recoil spectra.  With ExoDDDM, limits from XENON100 are further weakened due to the narrow width which keeps all events below the single photoelectron threshold, unlike for ExoDM where Poisson fluctuations can push some events above threshold.}
  \label{fig:spectra}
\end{figure}

\begin{figure}[t!]
  \centering
 \includegraphics[height=0.38\textwidth]{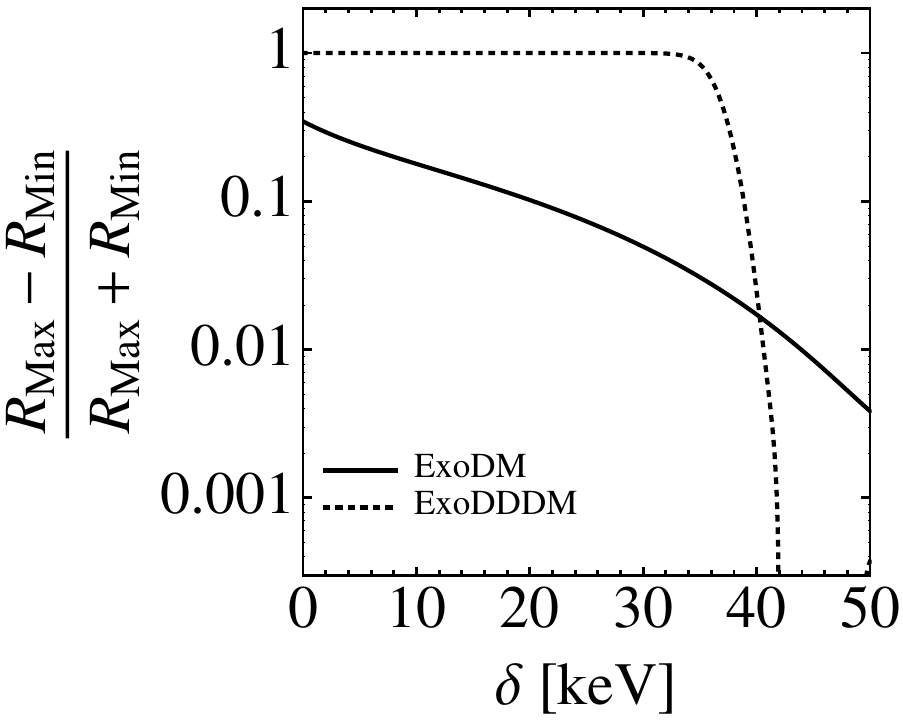}
  \caption{The ratio of modulated to unmodulated scattering rates in the CDMS-Si detector for $M=5.5$ GeV as a function of the exothermic splitting for ExoDM and ExoDDDM.  In both cases the introduction of an exothermic splitting leads to a reduction in the signal modulation.  Below $\delta \sim 32$ keV there is no signal for ExoDDDM since all scattering events are below threshold, hence the ratio is unity, but it should be kept in mind that the signal is vanishing in this region.}
  \label{fig:modamp}
\end{figure}

Exothermic scattering also leads to interesting spectral features.  In \Fig{fig:spectra} we show sample nuclear recoil spectra for ExoDM and ExoDDDM.  A feature common to both scenarios is that the recoil spectrum vanishes at low energies and is peaked at an energy which depends on the parameters of the model and the nucleus in question.  This is quite distinct from a standard elastically scattering DM spectrum which typically shows falling exponential behavior.  As direct detection experiments push to lower recoil energy thresholds this distinction will become increasingly important as the exponential rise at low energies for light (ordinary elastically-scattering) DM can be constrained more efficiently, whereas constraints on exothermic DM will depend instead on the typical exposure at the expected peak of the spectrum along with the width of the spectrum, which is determined by the velocity distribution.

There are also interesting implications for the annual modulation of the signal.  Since the exothermic splitting allows a detector to sample more of the DM velocity distribution than in the elastic scattering case, a smaller fraction of the total events arise from DM particles in the tail of the velocity distribution.  An implication of this is that the amplitude of the annual modulation signal is reduced relative to the unmodulated signal, which is a generic signature for exothermic DM scattering.  In \Fig{fig:modamp} we show the reduction in the modulation amplitude with increasing exothermic splitting for ExoDM and ExoDDDM scattering in the CDMS-Si detector.

This concludes the discussion of features arising due to exothermic scattering.  We now consider the specific differences that arise from the distinctive phase space distribution of DDDM.

At the core of the DDDM proposal is the possibility that a subdominant component of DM may behave in some ways similarly to visible matter and have small typical velocities so that it may also have collapsed within galaxies such as our own, forming structures similar to the galactic disk of visible matter with its higher density.  Without detailed numerical simulations we cannot make definite predictions for the properties of such a disk of DDDM, but we can use the visible baryonic disk and the discussion in \Refs{Fan:2013yva,Fan:2013tia} as guidance to estimate the phase space properties of the DDDM.   The quantities relevant to direct detection experiments are the local DDDM density, $\rho_0$, the velocity dispersion of the DDDM, $\tilde{v}$, and the relative velocity between the galactic disk and the visible baryonic disk, $v_{\text{rel}}$.

Due to uncertainties in the calculation of the dark disk thickness we treat the local DDDM density as a free parameter (estimates are discussed in \cite{Fan:2013yva,Fan:2013tia}).  We will plot results for an assumed local density of $0.3 \text{ GeV/cm}^3$ since this local density is possible for DDDM and it allows straightforward comparison between the required cross-sections for DDDM and standard halo DM.  However it should be kept in mind that the density may be higher or lower depending on the specific model under consideration.

In \cite{Fan:2013yva} a specific model of DDDM was constructed and the velocity dispersion estimated.  The particular value depends on the dark force coupling strength and the mass of the light states within the dark sector, so a broad range of values are possible.  We can estimate a reasonable value by considering a specific benchmark point discussed in \cite{Fan:2013yva}, where for a DM mass $M\approx 10$ GeV, the DDDM velocity dispersion is $\tilde{v} \approx 10^{-4} \text{ c}$.  Unless otherwise stated we choose a value $\tilde{v} = 25 \text{ km/s}$, and keep in mind that increasing or reducing $\tilde{v}$ from this value will increase or reduce the width of the recoil spectrum.

Finally we must consider the relative velocity of the visible baryonic and DDDM disks.  If a disk of DDDM exists in our galaxy then it may or may not lie in the plane of the visible baryonic disk.  If the two disks are not aligned then direct detection signals are unlikely unless the line of intersection of the two disks contains our solar system.\footnote{Such a scenario could lead to interesting signatures due to potentially large relative velocities between DDDM and the Earth.  Combined with the small velocity dispersion this scenario would look very much like a DM stream from the perspective of direct detection experiments.  We do not consider this scenario further here.}  Hence we consider only the possibility that both disks lie in the same plane.  In this case, if the dynamics of gravitational collapse and cooling were similar, it is likely that both disks have similar rotational velocities, and small relative velocities.  For this reason we will choose a benchmark relative velocity of $v_{\text{rel}}= 0 \text{ km}/\text{s}$, although different relative velocities are possible.\footnote{We will discuss the effect of varying the relative disk velocity in \Sec{sec:varying}.}  This completes the parameter specification required to begin extracting the qualitative features of ExoDDDM scattering, which we now outline.

\begin{figure}[t]
  \centering
\includegraphics[width=0.45\textwidth]{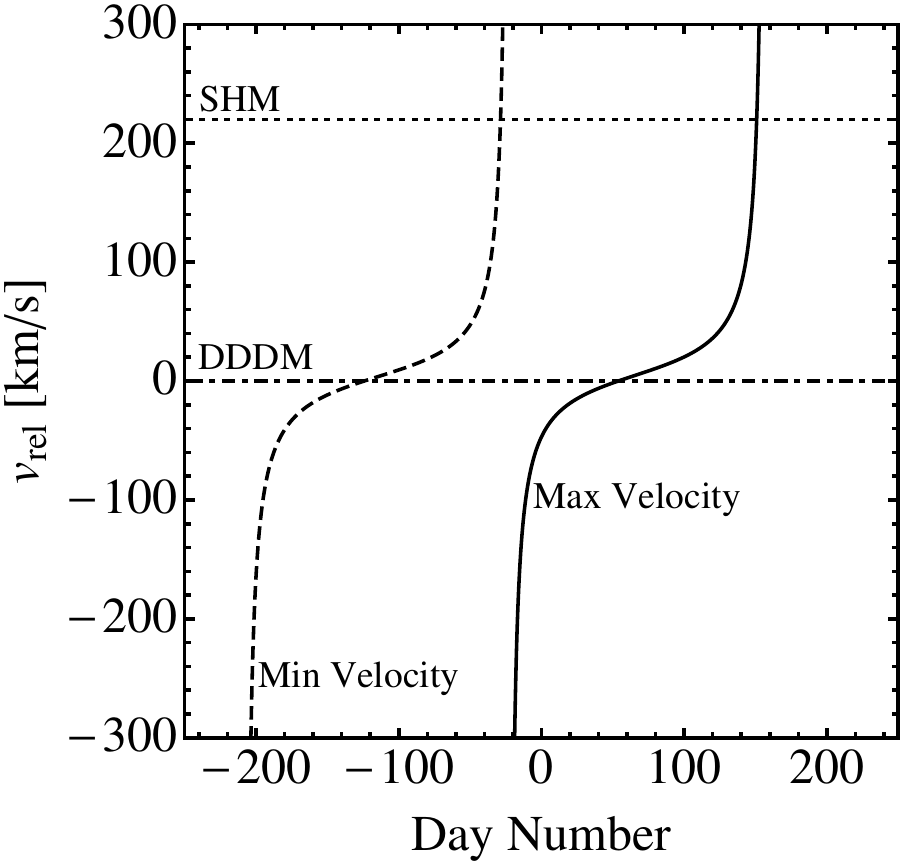} \hspace{0.2in}
  \caption{The relationship between the rotational velocity of DDDM relative to the visible baryonic disk, and the date on which the maximum and minimum Earth-DDDM relative velocity occurs.  The day number refers to the numbers of days relative to January $1^{\text{st}}$ 2000.  For standard halo DM the maximum relative velocity occurs around day $140$, due to a relative rotational velocity of $v_{\text{rel}} \approx220 \text{ km/s}$.  However for smaller relative velocities the peak velocity can occur much earlier in the year, leading to a significant shift in the phase of the annual modulation. The phase of the signal is more sensitive to the relative rotational velocity at small values.  Due to the decreased modulation amplitude signature of exoDDDM, a large number of events would be required to observe this phase.}
  \label{fig:phase}
\end{figure}

\begin{itemize}
\item  Narrow recoil spectra:  For a standard halo the DM velocity dispersion is expected to be as large as the visible baryonic disk rotational velocity, typically $\mathcal{O} (220) \text{ km/s}$, and this leads to a broad recoil energy spectrum, even for ExoDM.  On the contrary, for ExoDDDM, as the velocity dispersion is expected to be small, the recoil spectrum can be very narrow.  This feature is shown in \Fig{fig:spectra} where the typical ExoDM spectrum is much broader than the ExoDDDM spectrum.  With a large number of events a narrow recoil spectrum would be smoking gun for a small DM velocity dispersion and ExoDDDM scattering.

\item  Out-of-phase signal modulation:  For DM in a standard halo the relative velocity of the DM and the Earth is dominated by the large rotational velocity of the visible baryonic disk relative to the DM halo, which is $v_{\text{rel}} \approx \mathcal{O} (220) \text{ km/s}$.  Hence, once one also accounts for the peculiar velocity of the Sun and the orbit of the Earth, the dates of maximum and minimum relative velocity are predetermined, leading to clear predictions for the phase of the annual modulation signal.  For DDDM the relative disk-halo velocity is expected to be smaller, so the extremal velocities and scattering rates can occur on different dates.  For DDDM co-rotating with the visible baryonic disk the dates of maximum and minimum relative DM velocity are offset by $\mathcal{O}(100)$ days compared to standard halo DM, and small changes in the relative rotational velocity near $v_{\text{rel}} \approx 0 \text{ km/s}$ can lead to significant changes in the phase of the annual modulation, as shown in \Fig{fig:phase}.  Thus, if an annual modulation were observed with an unexpected phase this could point towards DDDM scattering.  However it should be kept in mind that a complementary signature of ExoDDDM is that the modulation amplitude can be greatly suppressed (see \Fig{fig:modamp}), and such a modified-phase signature would occur only if the splitting $\delta$ is not so great as to completely suppress any modulation.
\end{itemize}

With all of these features in mind we can now make a simple qualitative estimate of the recoil spectra for ExoDDDM scattering.  Since the nuclear recoil spectrum is narrow, and the DM is scattering essentially at rest, we can compare between different target nuclei by simply considering the recoil energy for scattering at rest
\be
E_R (\mb{v} = 0) =  \frac{\delta M_\chi}{M_N+M_\chi} ~~.
\label{eq:rest}
\ee
The true recoil spectrum, after integrating over DM phase space, will be narrow and peaked at characteristic energies given by \Eq{eq:rest}.

These qualitative features apply to the broad scenario of ExoDDDM and make it clear that optimal ExoDDDM search strategies require detectors with light nuclei, good energy resolution, and, in the event of DM discovery, an integrated exposure great enough to observe the phase of the annual modulation, although we note that the latter could be very challenging.

\section{An ExoDDDM Interpretation of the CDMS-Si Events}
\label{sec:quantitative}
For more than a decade the steady progress in DM direct detection technology has been punctuated by occasional experimental anomalies -- in particular, the long-standing DAMA annual modulation \cite{Bernabei:2010mq}, followed by the CoGeNT excess and modulation \cite{Aalseth:2011wp}, CRESST-II excess \cite{Angloher:2011uu}, and now the CDMS-Si excess \cite{Agnese:2013rvf}.  In most cases the anomalous scattering events arise near the experimental low-energy threshold, leading to tentative excitement over light DM interpretations, which is accordingly tempered by concerns over the significance of the signal and the tension between the simplest DM interpretations and null search results.

The differing targets and detection techniques of each experiment make a truly model-independent comparison among the various experiments difficult.  However powerful halo-independent strategies have been developed \cite{Fox:2010bu,Fox:2010bz,Frandsen:2011gi,Gondolo:2012rs,HerreroGarcia:2012fu,Bozorgnia:2013hsa} and have made it clear that elastically scattering light DM interpretations of all anomalies are now in tension with results from the XENON experiment \cite{Frandsen:2013cna,DelNobile:2013cta}.  However, this tension doesn't necessarily apply to all dark matter models and can be avoided if the microphysics is different.\footnote{For a thorough discussion of the elastic scattering case considering a potential resolution of the tension based on different XENON100 efficiencies see \cite{Hooper:2013cwa}.}  We know little about the dark sector, and are in no position to decide which possibilities are and are not reasonable, so any interesting modifications of DM scattering should be considered seriously.  Allowing for such modifications may then alleviate tensions between particular DM hints and exclusions.\footnote{In many scenarios where the scattering dynamics are altered, the local DM phase space distribution can become an important factor in determining scattering rates at different experiments and the significant uncertainties therein must always be considered, particularly when the disagreement between different experimental results is marginal \cite{Fairbairn:2008gz,MarchRussell:2008dy,Fox:2010bu,Fox:2010bz,McCabe:2010zh,McCabe:2011sr,Frandsen:2011gi,Green:2011bv,Fairbairn:2012zs,Gondolo:2012rs,HerreroGarcia:2012fu,Bozorgnia:2013hsa}.}

The three candidate scattering events in the CDMS-Si experiment, corresponding to a $\sim 3 \sigma$ excess, recently announced by the CDMS collaboration \cite{Agnese:2013rvf}, are intriguing for a number of reasons.  The experiment can distinguish nuclear and electromagnetic scattering events very well and the three observed events have properties consistent with nuclear recoils, as expected for DM scattering.  Furthermore, due to extensive calibration the background is well understood and explanations of the events based on backgrounds are not forthcoming.  For these reasons, and in light of the significant tension between limits from the XENON experiments and standard light DM interpretations of the CDMS-Si events, it is interesting to consider whether ExoDDDM may offer an explanation of these events in better agreement with the null search results.  Even if this signal does not survive, we will learn about ways to distinguish among different categories of DM and what types of measurements will prove important.  We delay specific details of the analysis methods to \App{sec:details} and focus here on the results.

\begin{figure}[t]
  \centering
  \includegraphics[height=0.39\textwidth]{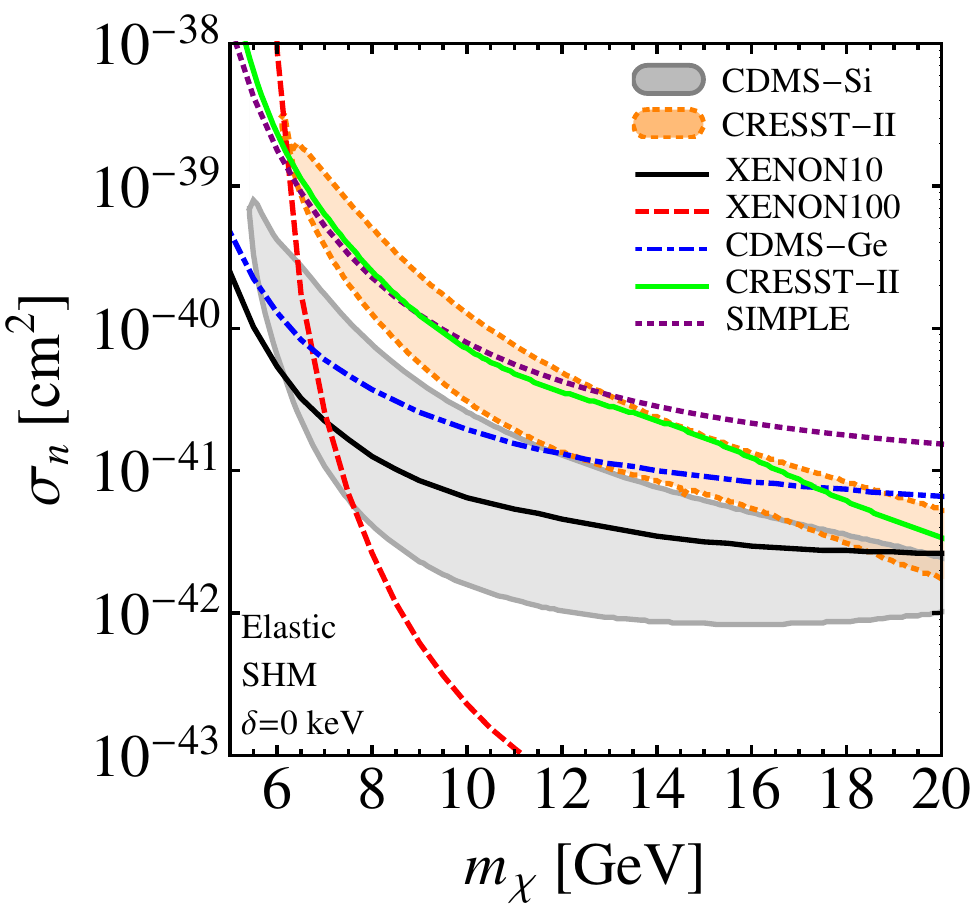} \hspace{0.2in}  \includegraphics[height=0.39\textwidth]{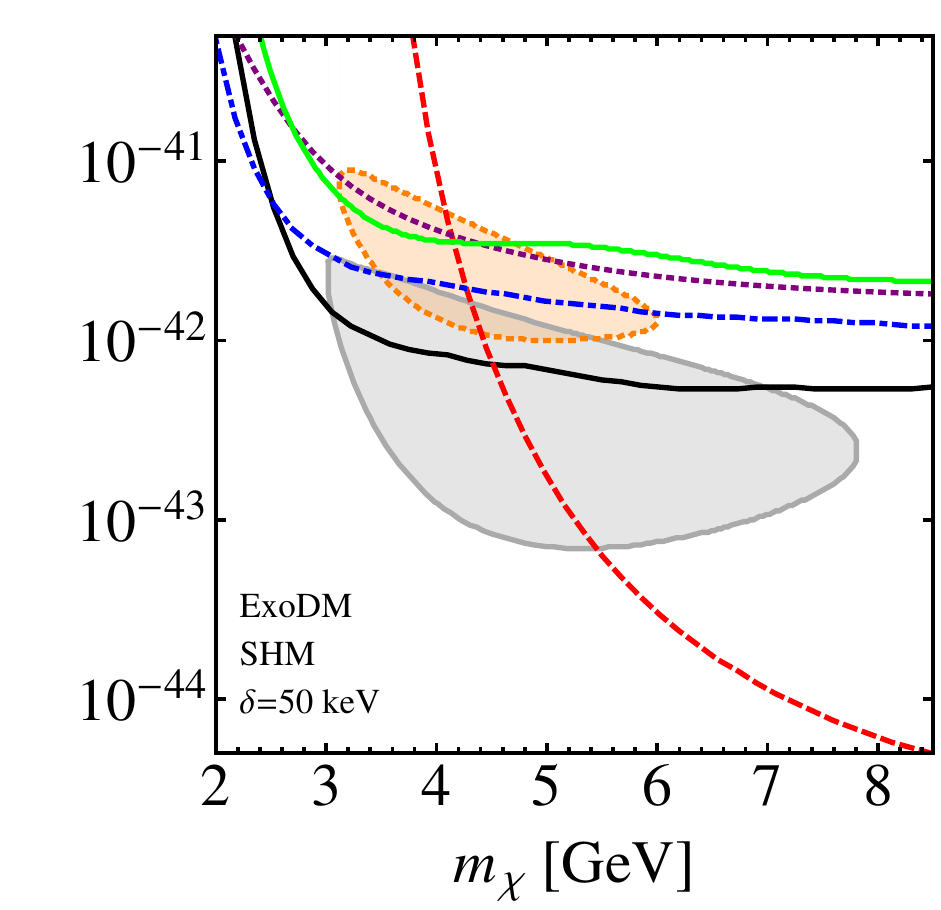} \\ \vspace{0.2in}
   \includegraphics[height=0.39\textwidth]{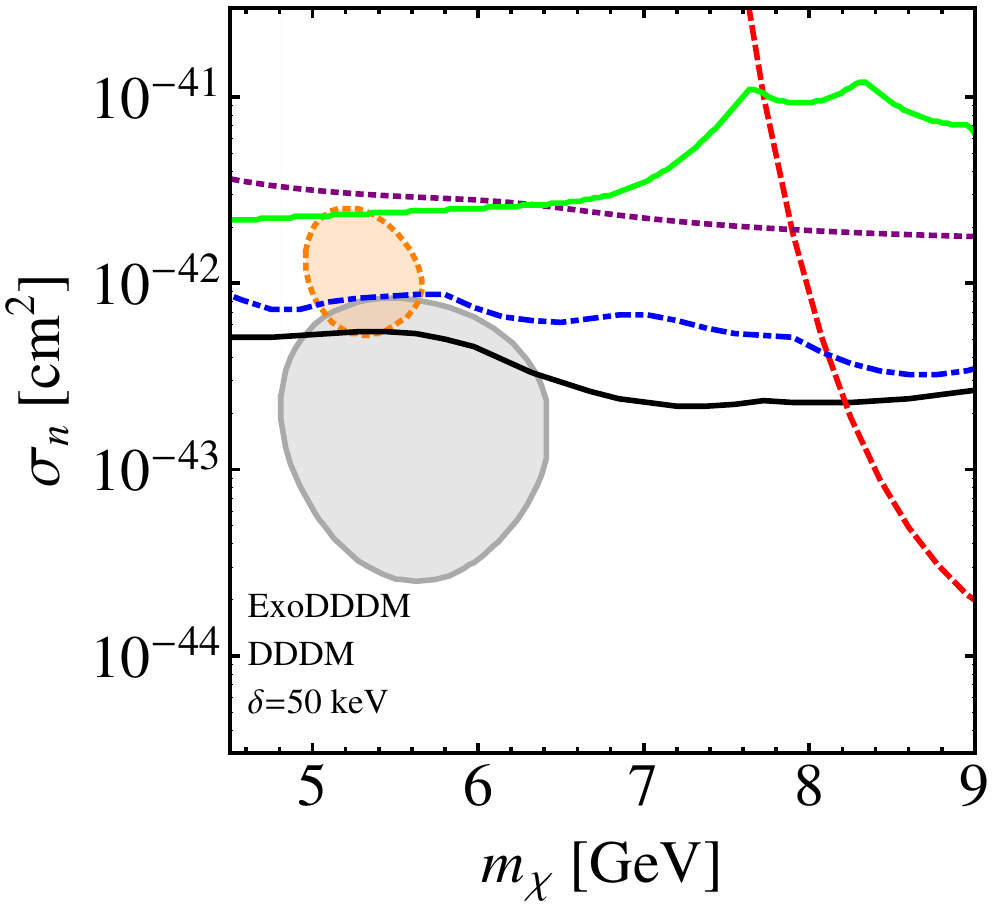} \hspace{0.2in}  \includegraphics[height=0.39\textwidth]{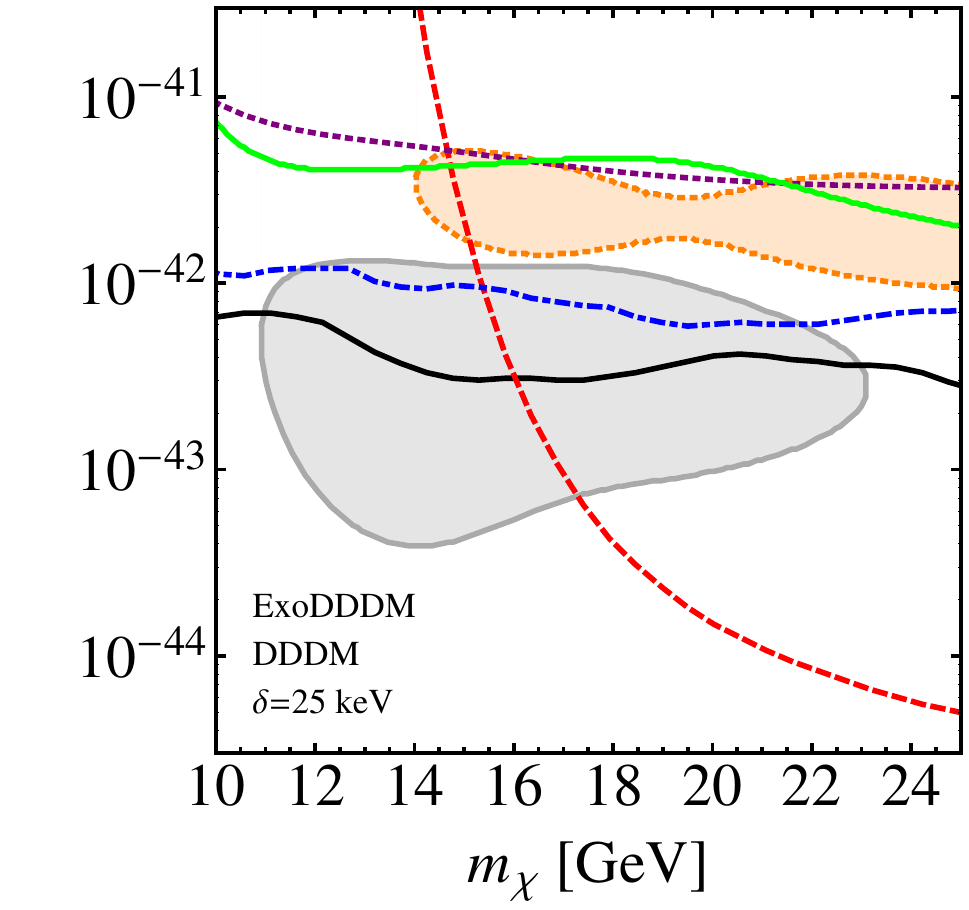}
  \caption{$90\%$ best-fit regions (CDMS-Si shaded gray and CRESST-II shaded orange) and $90\%$ exclusion limits (XENON10 solid black, XENON100 dashed red, CDMS-Ge dot-dashed blue, CRESST-II low threshold analysis solid green and SIMPLE in dotted purple). Elastic and exothermic scattering of standard halo DM are shown in the upper panels, and ExoDDDM below.  Elastic scattering of light DM gives a good fit to the CDMS-Si events, although there is significant tension with null results.  ExoDM reduces the tension and opens up additional parameter space consistent with CDMS-Si and limits from the null search results \cite{Frandsen:2013cna}.  ExoDDDM scattering allows for a CDMS-Si interpretation with heavier DM mass (lower right).  For lighter ExoDDDM (lower left), the majority of the favored parameter space is consistent with the strongest bounds and the DM mass favored in asymmetric DM models.}
  \label{fig:fits}
\end{figure}

We consider limits from all experiments with strong sensitivity to spin-independent light DM scattering.  The strongest bounds typically come from the S2-only XENON10 analysis \cite{Angle:2011th} and from XENON100 \cite{Aprile:2012nq}.  We also derive limits from the CDMS-Ge dedicated low-threshold analysis \cite{Ahmed:2010wy}.  However we do not consider the CDMS-Ge annual modulation analysis \cite{Ahmed:2012vq} as the low-energy threshold of $5$ keV weakens constraints on ExoDDDM in this case, as evident from \Fig{fig:spectra}.  Also, as the modulation amplitude is suppressed for any exothermic model of DM, limits from modulation studies will be further weakened.  For the sake of thoroughness we also consider limits from the SIMPLE experiment \cite{Felizardo:2011uw} and the analysis of CRESST-II commissioning run data which includes oxygen recoils to push sensitivity to lower DM masses \cite{Brown:2011dp}.  As well as the $90\%$ best-fit region for the CDMS-Si events \cite{Agnese:2013rvf}, we also consider $90\%$ best-fit regions for the CRESST-II excess of events \cite{Angloher:2011uu}, however it should be noted that this excess, or at least a significant portion of it, may be accounted for with additional background sources \cite{Kuzniak:2012zm}.\footnote{The DAMA annual modulation \cite{Bernabei:2010mq} and the CoGeNT modulation \cite{Aalseth:2011wp} will be discussed in \Sec{sec:multiple}.}

In \Fig{fig:fits} we show best-fit contours for the CDMS-Si and CRESST-II anomalies alongside experimental limits for elastically scattering standard halo DM, ExoDM and ExoDDDM.  While an interpretation of the CDMS-Si and CRESST-II anomalies based on elastic scattering of light DM is in tension with XENON10 and XENON100 under the assumption of the standard halo model, an ExoDM interpretation considerably alleviates this tension \cite{Frandsen:2013cna}.  Intriguingly, if the scattering were instead due to DDDM particles then there is virtually no constraint on an ExoDDDM interpretation of the CDMS-Si excess for DM with mass in the region predicted by many asymmetric DM models \cite{asymm}.

The almost total consistency between an ExoDDDM explanation of the CDMS-Si excess and the XENON limits results from the combination of exothermic scattering kinematics with the DDDM phase space distribution, as described in \Sec{sec:pheno}.  The CDMS-Si events are at energies of 8.2, 9.5, and 12.3 keV and the energy resolution of the detector is 0.5 keV.  By considering the silicon recoil spectrum shown in \Fig{fig:spectra} for the 5.5 GeV ExoDDDM scenario it is clear that the 8.2 and 9.5 keV events are accommodated well by the ExoDDDM scattering.  For the DDDM velocity distribution chosen here, the 12.3 keV event does not originate from ExoDDDM scattering.  However, this does not lead to a bad fit for ExoDDDM scattering since $\mathcal{O}(0.7)$ events are expected from background alone, and the 12.3 keV event is accommodated by this expected background rate.

If the CDMS-Si events really were due to ExoDDDM then this explanation could be verified in two ways.  First, since the recoil spectrum is much more peaked than for standard halo DM or ExoDM, further integrated exposure with a silicon detector should see events accumulate in the $\sim 9$ keV region, but not at lower or higher recoil energies.  The ExoDDDM interpretation of the CDMS-Si excess would also lead to a modification of the amplitude and phase of the annual modulation.  \Fig{fig:modamp} shows that for the parameters chosen the amplitude of the modulation would be very suppressed, which is a generic feature of exothermic DM scattering, and the modified phase of the modulation would be difficult to observe.

\subsection{Varying DDDM Phase Space Parameters}
\label{sec:varying}
As discussed in \Sec{sec:pheno} there exist uncertainties in the phase space distribution of the dark disk.  Thus it is interesting to consider if the ExoDDDM explanation of the CDMS-Si excess demonstrated in \Sec{sec:quantitative} is strongly influenced by changes in the disk parameters.  The two relevant parameters which could lead to changes in the agreement between the ExoDDDM CDMS-Si interpretation and limits from the XENON experiments are the relative velocity between the dark disk and the visible baryonic disk, $v_{\text{rel}}$, and the velocity dispersion of the dark disk $\tilde{v}$.

\begin{figure}[h]
  \centering
  \includegraphics[height=0.39\textwidth]{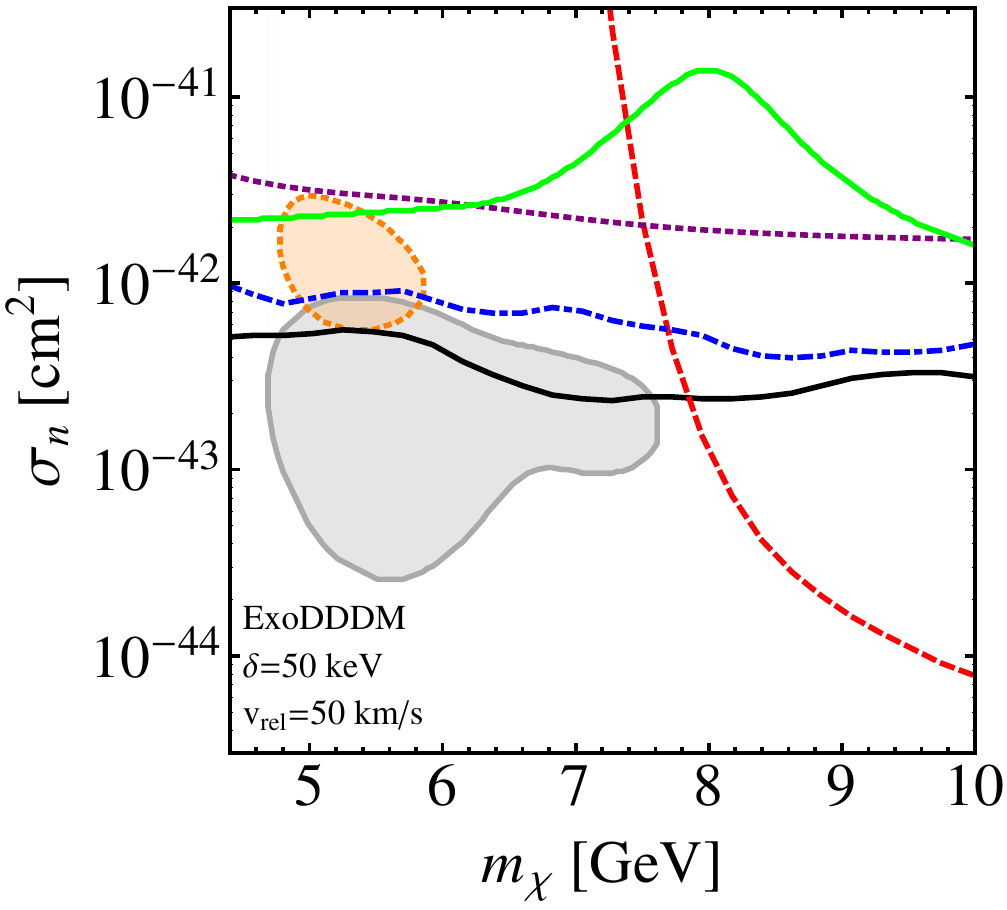} \hspace{0.2in}  \includegraphics[height=0.39\textwidth]{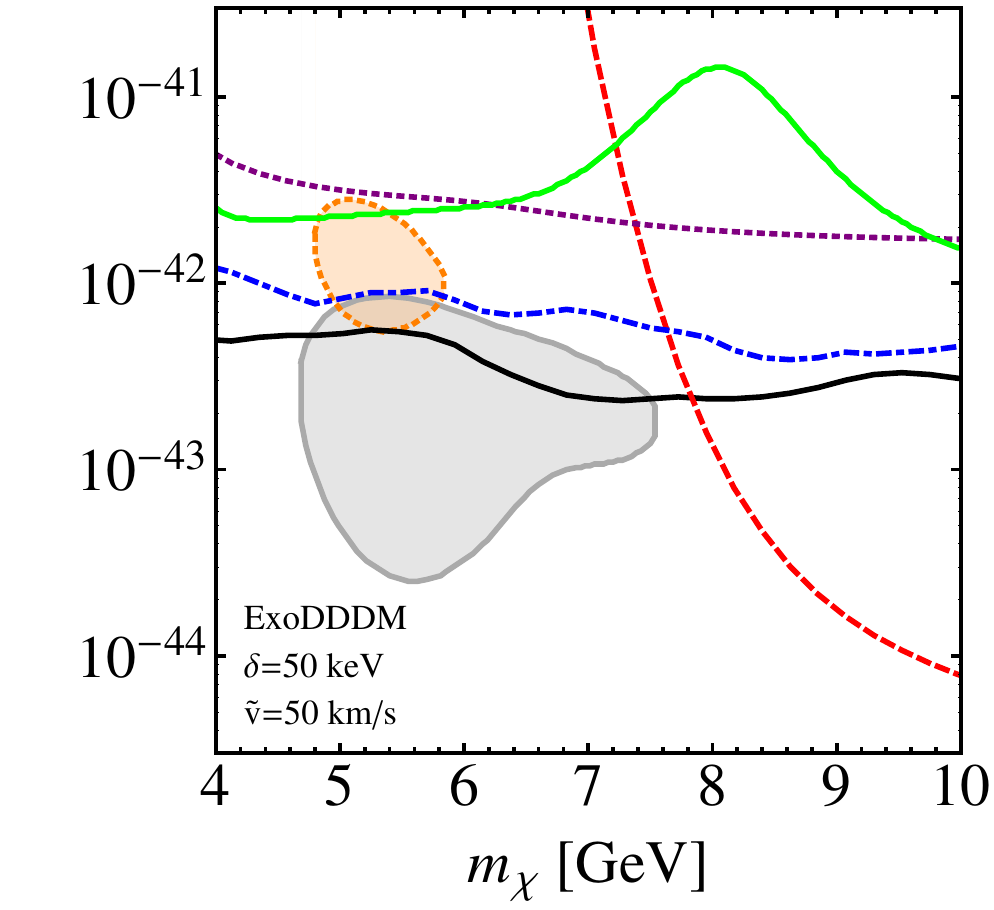}
  \caption{$90\%$ best-fit regions and exclusion limits for ExoDDDM (colors as in \Fig{fig:fits}) with varied dark disk phase space parameters.  On the left panel the dark disk rotational velocity is slower than the visible baryonic disk by $50$ km/s.  On the right panel the velocity dispersion of the dark disk is increased to $50$ km/s.  Both cases demonstrate that the consistency between an ExoDDDM explanation of the CDMS-Si events and exclusion limits from the XENON experiments is robust to uncertainties in the dark disk parameters.}
  \label{fig:vary}
\end{figure}

In the left panel of \Fig{fig:vary} we show the best-fit regions and exclusion limits for a relative rotational velocity which has been increased from a co-rotating disk $v_{\text{rel}} = 0$ km/s (\Fig{fig:fits}) to a disk which lags the visible baryonic disk rotation by $v_{\text{rel}} = 50$ km/s.  On the right panel of  \Fig{fig:vary} we show the impact of doubling the velocity dispersion from $\tilde{v} = 25$ km/s (\Fig{fig:fits}) to $\tilde{v} = 50$ km/s.  In both cases the consistency between the ExoDDDM interpretation of the CDMS-Si excess and limits from the XENON limits persists, demonstrating that the ExoDDDM interpretation of the CDMS-Si excess is robust to uncertainties in the dark disk phase space parameters.

\subsection{Looking Below Thresholds}
\label{sec:belowthresholds}
Further interesting ExoDDDM signatures arise when considering \emph{below} threshold events.\footnote{This was emphasized in \cite{NealTalk} and will be discussed in detail in \cite{Weineretal}.  We thank Patrick Fox and Neal Weiner for conversations concerning the importance of below-threshold events.}  Here, and in \cite{Aprile:2012nq,Frandsen:2013cna}, when calculating limits from XENON100 \cite{Aprile:2012nq} a low-energy threshold of $6.6$ keV, corresponding to three S1 photoelectrons, is employed.  This is shown in \Fig{fig:spectra} where one can see that in both ExoDM and ExoDDDM scenarios the vast majority of xenon scattering events lie below this threshold for the parameters chosen to fit the CDMS-Si excess.  For scattering energies below $3.4$ keV, also shown in \Fig{fig:spectra}, the detection efficiency is effectively zero since one expects less than one S1 photoelectron. Hence this represents a hard low energy threshold.  However it may be possible to consider a lower energy threshold, from $6.6$ keV down to $3.4$ keV if a reliable extrapolation of the efficiency could be performed.

In Fig.\ $2$ of \cite{Aprile:2012nq} no signal events are observed below the low-energy threshold of $6.6$ keV down to $3.4$ keV.  Hence if one could reliably extend the low energy threshold into this region then, by comparison with \Fig{fig:spectra}, the XENON100 limits could become significantly more constraining on light elastically scattering DM and ExoDM interpretations of the CDMS-Si excess.  On the other hand, due to the narrow spectrum for ExoDDDM the xenon scattering events all lie below the single photoelectron threshold, and it is unlikely that ExoDDDM interpretations of the CDMS-Si events would become significantly constrained in this case.  Thus an improved understanding of the low energy efficiency of the XENON100 experiment could significantly narrow the range of models which could explain the CDMS-Si events.

We also show the low energy threshold for the CDMS-Si analysis in \Fig{fig:spectra}.  For ExoDM increased integrated exposure with a silicon detector would lead to additional events below the $7$ keV threshold set by CDMS \cite{Agnese:2013rvf}.  However for ExoDDDM the spectrum is considerably more narrow, and additional events would not be expected below the $7$ keV threshold.

A push to lower thresholds with both xenon and silicon-based detectors could illuminate DM interpretations of the CDMS-Si events, and further constrain, or support, ExoDM and ExoDDDM scenarios.

\subsection{Fitting Multiple Anomalies}
\label{sec:multiple}
In \Fig{fig:fits} we have shown best-fit regions for CDMS-Si and CRESST-II.  We have not shown best-fit regions for the DAMA and the CoGeNT modulation signals.  Interpretations of these anomalies are in tension with limits from the XENON experiments for elastic scattering.  ExoDM and ExoDDDM also both give a bad fit to these anomalies for parameters which fit the CDMS-Si events, as we now demonstrate.

We can use \Eq{eq:rest} to estimate the typical exothermic splitting, $\delta$, required to fit a particular anomaly by requiring that the typical recoil energy lies within the energy range of anomalous nuclear recoils at each detector.  We estimate these energy ranges in \Tab{tab:energies} where we have used the sodium quenching factor $q_{N_a} = 0.3$ for DAMA.

\begin{table}[t!!]
\centering
\begin{tabular}{c l}
Typical $E_R$ & Experiment  \\ \hline
$10.0 \lesssim E_O \lesssim 22.0$ keV & (CRESST-II) \\
$8.2 \lesssim E_{Si} \lesssim 12.3$ keV & (CDMS-Si) \\
$7.5 \lesssim E_{N_a} \lesssim 12.5$ keV & (DAMA) \\
$0.5 \lesssim E_{Ge} \lesssim 2.0$ keV & (CoGeNT)
\end{tabular}
\caption{Typical nuclear recoil energies of anomalous events at various experiments.}
\label{tab:energies}
\end{table}

\begin{figure}[h!]
  \centering
  \includegraphics[height=0.43\textwidth]{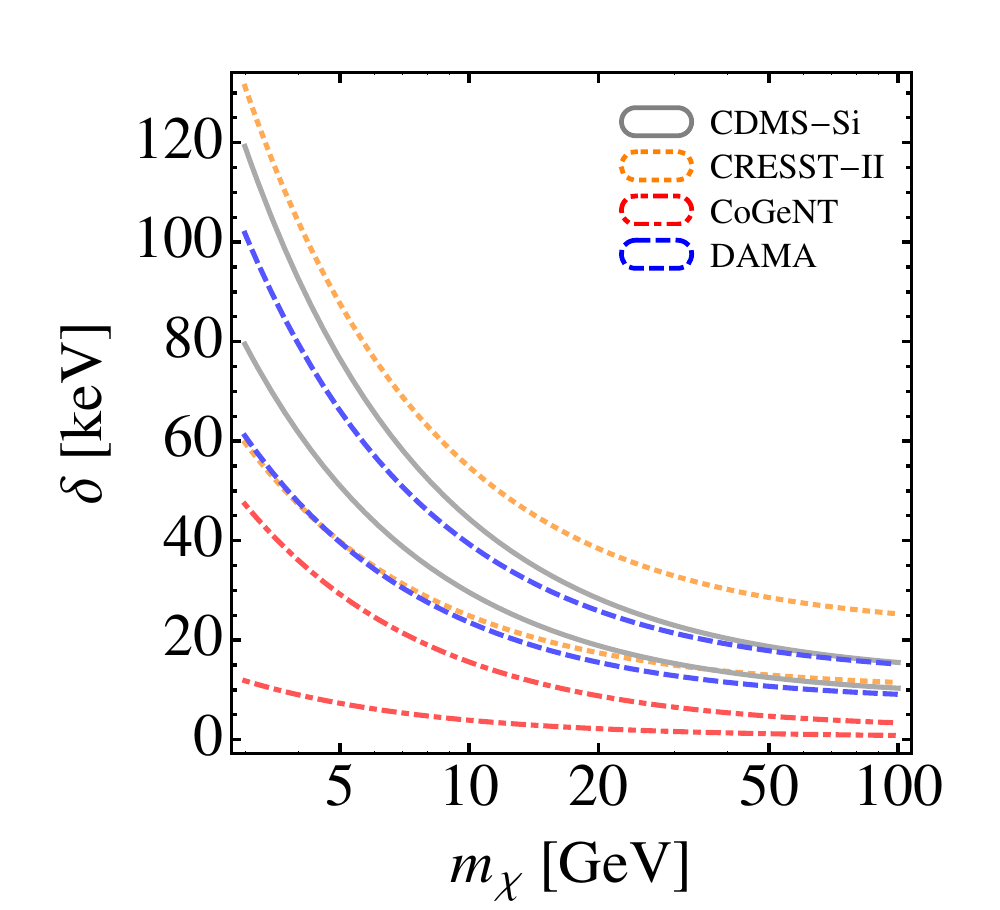} \hspace{0.2in} \includegraphics[height=0.41\textwidth]{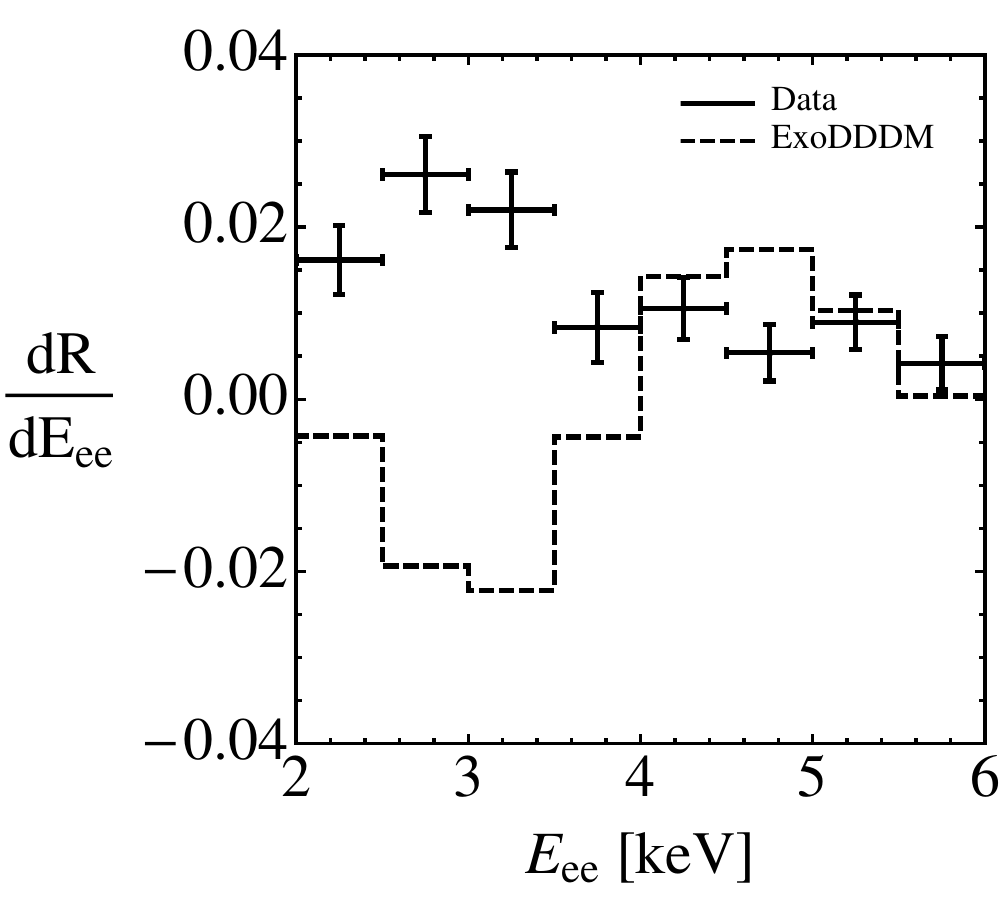}
  \caption{The required exothermic splitting $\delta$ required to give nuclear recoil energies in the typical range required by the various anomalies (left panel).  Required values for a given experiment lie between the two corresponding contours.  Splittings required for CoGeNT are inconsistent with those required for CRESST-II, CDMS-Si and unquenched DAMA.  We also show the modulation spectrum at unquenched DAMA for $\delta = 50$ keV, $M=5.5$ GeV, and $\sigma_n = 10^{-40} \text{cm}^2$ (right panel).  We have increased the relative rotational velocity to $v_{rel} = 100 \text{ km/s}$ to give dates of maximum and minimum relative velocity closer to the dates of maximum and minimum event rates in DAMA, however the spectrum is in anti-phase with the data.}
  \label{fig:anomalies}
\end{figure}

In the left panel of \Fig{fig:anomalies} we find the required values of $\delta$ for these energy ranges, and show the preferred range of $\delta$ between corresponding contours.  It is immediately clear that the CDMS-Si and CoGeNT modulation regions do not overlap, making a consistent ExoDDDM explanation for the CoGeNT modulation and CDMS-Si signals unlikely.\footnote{If the channelling effect \cite{Bozorgnia:2010xy} is included for DAMA then the required values of $\delta$ for DAMA are shifted downwards by a factor of $0.3$, and a common ExoDM origin for the DAMA and CoGeNT modulations can be found, as demonstrated in \cite{Graham:2010ca}.  However this would not move the CRESST-II and CDMS-Si regions into agreement with CoGeNT, and it is clear that a common ExoDDDM interpretation of CRESST-II, CDMS-Si and CoGeNT is unlikely.
}

For a relative rotational velocity between the dark disk and visible baryonic disk of $v_{rel} = 0 \text{ km/s}$ \Fig{fig:phase} shows that the expected phase of the annual modulation would be incorrect for fitting DAMA.  However, increasing the rotational velocity to $v_{rel} = \mathcal{O} (\gtrsim 60) \text{ km/s}$ would bring the dates of maximum and minimum relative DM velocities in line with typical expectations for standard halo DM scattering.  Thus the date of maximum DM velocity for DAMA can be broadly compatible with an ExoDDDM explanation of CRESST-II and CDMS-Si.  However, due to the exothermic scattering, the amplitude of the modulation predicted at DAMA has the wrong sign.  This is shown for typical masses and splittings, but an enhanced cross-section on the right panel of \Fig{fig:anomalies}, demonstrating that parameter choices which give a good fit to CDMS-Si typically give a bad fit to the DAMA modulation as the sign of the modulation is wrong and the cross-section would be too low to explain the amplitude of the annual modulation.  Hence a common ExoDDDM explanation of all four direct detection anomalies is unlikely.  However, a consistent ExoDDDM interpretation of the CDMS-Si and CRESST-II events is possible.  %Furthermore the cross-section would be too low to explain the large modulated signal.

\begin{figure}[t!]
  \centering
  \includegraphics[height=0.43\textwidth]{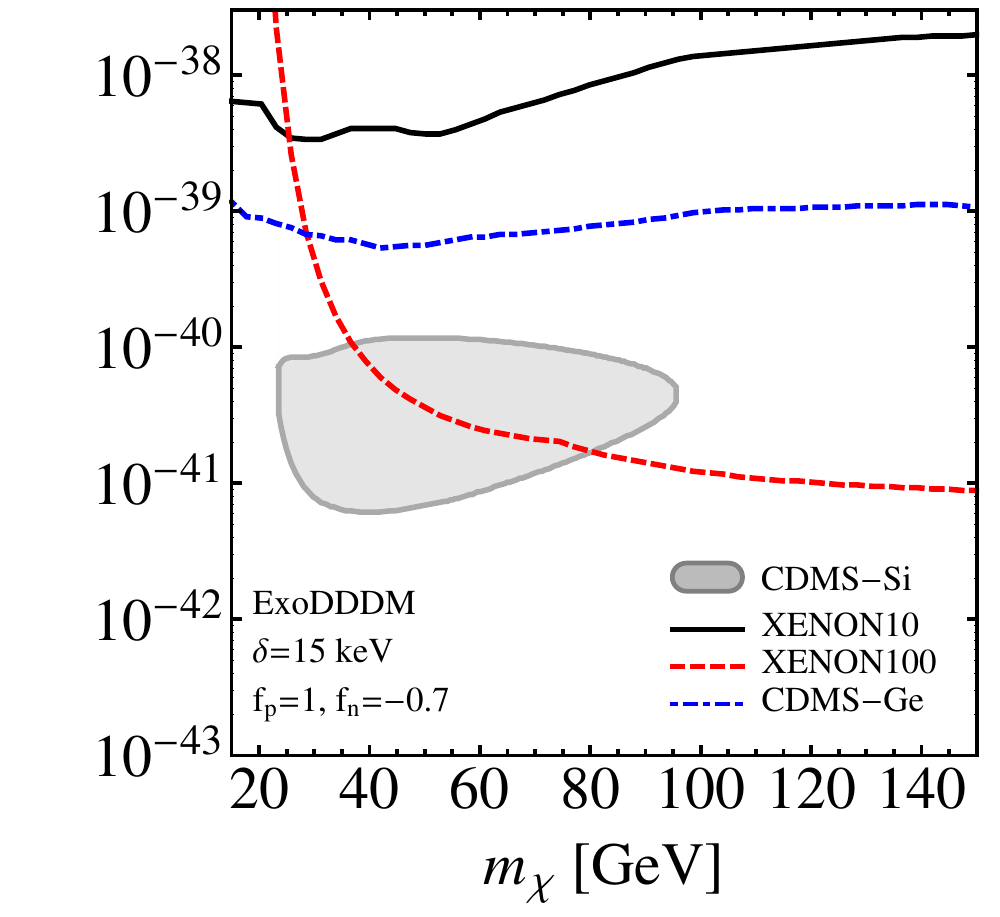}
  \caption{Bounds on a heavy ExoDDDM interpretation of the CDMS-Si events when DM couplings to protons and neutrons are tuned to suppress bounds from XENON10 and XENON100.  This introduces new higher-mass interpretations of the CDMS-Si events which are consistent with exclusion limits.}
  \label{fig:isospin}
\end{figure}

\subsection{Introducing Isospin-Dependent Couplings}
All results presented thus far have been under the assumption that the DM couples equally to protons and neutrons, i.e.\ $f_n=f_p$.  However, if this is not the case then the relative scattering rates at different detectors are modified depending on the number of protons and neutrons within the nucleus \cite{Kurylov:2003ra,Giuliani:2005my,Chang:2010yk,Feng:2011vu,Feng:2013vod}.  For silicon at CDMS-Si and oxygen at CRESST-II we have $(A-Z)=Z$. Hence the scattering rates scale as $\propto (f_n+f_p)^2$ and isospin-dependent couplings will not improve the agreement between the best-fit regions for both experiments.  However, isospin-dependent couplings can significantly weaken limits from XENON10 and XENON100 for certain values of these couplings.

For ExoDDDM with $M\sim 5$ GeV there is already almost complete agreement between the best-fit regions and XENON limits and there is no motivation to include isospin-dependent couplings in this case.  However, if ExoDDDM scattering and isospin violation are combined a DM interpretation of the CDMS-Si events for larger DM masses is allowed.  For small splittings $\delta \sim 15$ keV a fit of ExoDDDM to CDMS-Si prefers greater DM masses.  We calculate XENON limits using the isotope abundances of  \cite{Feng:2011vu}, and in \Fig{fig:isospin} show that, by allowing for isospin-dependent couplings to be tuned to suppress the sensitivity of XENON100, some regions with relatively large DM masses, $M\lesssim 80$ GeV can open up.

\section{Indirect Constraints}
\label{sec:colliderindirect}
Here we estimate and discuss the indirect bounds from colliders and solar capture on broad classes of models which provide an ExoDDDM interpretation of the CDMS-Si events.  We find that collider production and solar capture do not lead to strong constraints although solar capture can be interesting in the future.

Although the DM is neutral and cannot be observed directly, it could in principle show up indirectly at colliders as a missing energy signature: $\cancel{E}_T+X$, where $X$ could be a mono-photon, mono-Z, or mono-gluon.  The cross-sections for these processes depend on the particular SM operator coupling the DM to SM states, the size of the coupling, and also the mass of the states which mediate the interactions.  For heavy mediators the limits on spin-independent couplings do not typically constrain DM-nucleon cross-sections below $\sigma_n \sim 10^{-40} \text{ cm}^2$ \cite{Primulando:2012xla}, and these limits are further weakened if the mediator is light.  Thus collider bounds place no strong constraint on the best-fit parameters considered here.  However, if the local DDDM density were lower a larger cross-section would be required to explain the CDMS-Si events, the implied DM-visible sector interaction strength may be greater, and collider signatures may be possible in the near future.

If the DM is symmetric and can annihilate to SM final states additional indirect constraints arise from neutrino detectors.  If $u$ is the speed of a DM particle, determined by the DM velocity distribution, and $v(r)$ is the speed which a particle picks up after falling into the Sun at a radius $r$ from the centre, then at this point a DM particle which has fallen into the Sun will have a total speed $w(r) = \sqrt{u^2+v^2 (r)}$.  At this point it could scatter on any of the nuclei contained within the Sun and if the new speed after scattering, which we denote $\widetilde{w}(r)$, satisfies $\widetilde{w}(r) < v (r)$ it will become gravitationally bound.  Over time it may continue to scatter, lose kinetic energy, and fall into the centre of the Sun, eventually annihilating with other captured DM particles.  If the DM annihilation cross-section is large enough then the annihilation rate becomes limited only by the capture rate, and the total rate of DM annihilation in the Sun is determined solely from the total capture rate.  If the annihilation final states contain SM states then decays or re-scattering of these states can produce neutrinos which propagate to the Earth and can be detected.  This leads to indirect constraints on DM from neutrino observatories such as Super-Kamiokande \cite{Desai:2004pq} and IceCube \cite{Aartsen:2012kia}.  The IceCube limits are typically most constraining for DM masses $M \gtrsim 10$'s GeV, and limits from Super-Kamiokande are typically strongest for lighter DM, in the region of $M \sim 5$ GeV.

For DM with non-standard kinematic properties, such as ``Inelastic DM'' \cite{TuckerSmith:2001hy} the dynamics of capture in the Sun \cite{Nussinov:2009ft,Menon:2009qj,Shu:2010ta} or other astrophysical bodies \cite{McCullough:2010ai} has been shown to exhibit significant departures from the standard elastically scattering case.  Therefore consideration of indirect limits from annihilation in the Sun, specifically for ExoDDDM, is deserving of its own complete study.  Here we will simply estimate the bounds.  If the scattering is exothermic then some portion of the exothermic energy released will go into giving the DM a small kick, sometimes increasing its kinetic energy.  Since this kick increases the speed, it reduces the fraction of particles which, after scattering, have speeds below the escape velocity and can in some cases suppress the solar capture of ExoDDDM.  Whether or not this suppression is important depends on the balance between the exothermic splitting energy $\delta$, and the kinetic energy picked up by falling into the Sun.  If the latter is dominant then the exothermic splitting will be unimportant.  If the former dominates then the suppression of capture can be non-negligible.

Once the first exothermic scattering has occurred the DM is in the lower of the two states.  To scatter again at tree-level will require inelastic scattering, and if the DM particle does not have sufficient kinetic energy this will be kinematically blocked.  However, the DM will also be capable of scattering elastically at one-loop and this cross-section is typically large enough to enable further scattering and energy loss of the DM particle.

\begin{figure}[t!]
  \centering
  \includegraphics[height=0.43\textwidth]{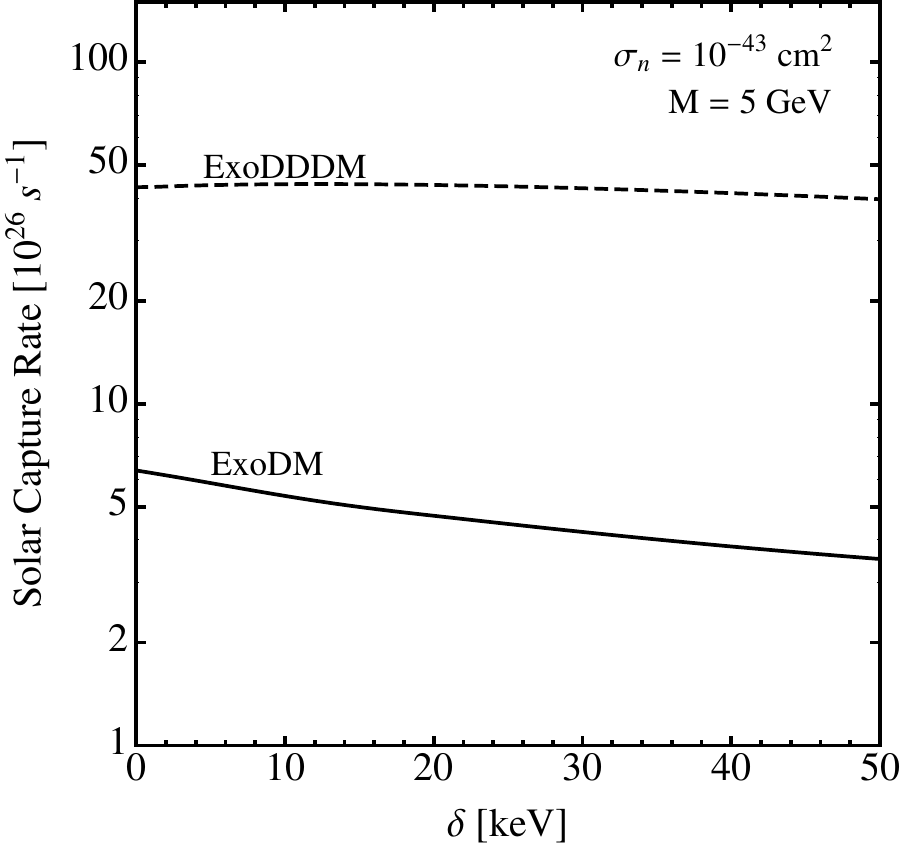}
  \caption{Solar capture rates for scattering on different nuclei for ExoDM (solid) and ExoDDDM (dashed).  The benchmark DM mass and cross-section considered here are $M=5$ GeV and $\sigma_n = 10^{-43} \text{ cm}^2$.  Due to the small relative velocity DDDM capture is typically enhanced relative to standard halo DM, and there is little variation due to the exothermic splitting for ExoDDDM, with some suppression when the splitting is increased in the ExoDM case.}
  \label{fig:capture}
\end{figure}

To estimate the solar capture rate we follow the calculation of \cite{Shu:2010ta} and use the Solar element abundances of \cite{Asplund:2009fu}.  The capture rates are shown in \Fig{fig:capture} for a choice of DM mass and cross-section which allows a fit to the CDMS-Si events.  In the elastic limit, which corresponds to the $\delta\rightarrow 0$ limit in \Fig{fig:capture}, one can see that DDDM capture is greatly enhanced over standard halo DM capture.  This results from the much lower velocity of DDDM compared to standard halo DM falling into the Sun.   If the initial speed of the DM is much larger than the escape velocity then, after scattering, only a small fraction of DM particles will have speeds less than the escape velocity, and only this small fraction will become captured.  However if the initial speed is comparable to, or smaller than the escape velocity then a larger fraction of scattered particles will have new speeds below the escape velocity, and the capture rate is subsequently enhanced.\footnote{This is also described in \cite{Gould:1987ir,Gould:1987ww}.}

In \Fig{fig:capture}, as the exothermic splitting is increased the capture rate is suppressed for ExoDM but not significantly for ExoDDDM.  For ExoDDDM the kinematics is entirely dominated by the kinetic energy picked up by falling into the Sun, and the exothermic splitting is largely irrelevant.  For ExoDM the initial speeds are greater and so a smaller fraction of scattered particles become captured.  The effect of the exothermic splitting ``kicking'' the DM back out of the Sun is more pronounced for this small fraction of events, and hence the splitting becomes more important for ExoDM.

To estimate solar capture limits we consider a recent analysis \cite{Kappl:2011kz} which studied spin-independent elastic scattering of light DM.  The most constrained final state is when the DM annihilates $100\%$ into neutrinos.  This final state is unlikely for concrete models, however we consider limits on this final state as a way of estimating the strongest bounds possible.  These limits exclude DM-nucleon cross-sections $\sigma_n \gtrsim 10^{-41} \text{ cm}^2$ for standard halo DM of mass $M \sim 5 \text{ GeV}$ which scatters elastically \cite{Kappl:2011kz}.  \Fig{fig:capture} shows that the capture rate for ExoDDDM with $M \sim 5 \text{ GeV}$ and $\delta \sim 50$ keV is enhanced relative to standard elastically scattering DM by a factor of $\sim 6$.\footnote{The enhancement follows almost entirely from the lower relative velocity of the DDDM.}  Hence we can estimate the scattering cross-section bound for ExoDDDM annihilation into neutrinos to be $\sigma_n \lesssim 1.7 \times 10^{-42} \text{ cm}^2$.  This is approaching, but does not exclude, the required typical cross-section for an explanation of the CDMS-Si excess, which is $\sigma_n \approx 10^{-43} \text{ cm}^2$.\footnote{As stated in \Sec{sec:pheno} there is a large uncertainty in the local DDDM density, and hence the cross-section required to explain the CDMS-Si events may vary greatly.  However the solar capture rate scales with the product of cross-section and DM density in the same way, hence variations in the DDDM density will not improve or degrade the limits from solar capture.}  Thus solar limits do not strongly constrain an ExoDDDM explanation of the CDMS-Si bounds, even if the most constrained final state is chosen.  However since the bounds are within two orders of magnitude it would be interesting to perform a more precise study which includes re-scattering effects in the approach to equilibrium.

Since the DDDM must have large self-interactions additional contributions to the capture rate can arise from DM self-capture.  However, as already stated, for any symmetric component of the DM the annihilation rate is typically faster than the capture rate. Hence the number of DM particles within the Sun, which must be scattered upon for self-capture, is kept low due to the efficient annihilation.  Thus for models of symmetric ExoDDDM this effect is unlikely to be important.

Alternatively, in models of ExoDDDM which contain an asymmetric component, such as the model presented in \Sec{sec:model}, not all of the DM annihilates.  In this case the self-capture of ExoDDDM can be important and the build up of ExoDDDM could lead to interesting effects on the inner dynamics of the Sun.  However at present any constraints are not strong \cite{Frandsen:2010yj,Taoso:2010tg,Kouvaris:2010jy,McDermott:2011jp,Iocco:2012wk}.  A complete calculation including both symmetric and asymmetric components, allowing for ExoDDDM capture and self-capture (including re-scattering effects) followed by the subsequent annihilation of the symmetric components into various final states is beyond the scope of this work.  However, we have demonstrated that strong constraints on ExoDDDM interpretations of the CDMS-Si excess are unlikely to arise from solar capture.  That said, it would be very interesting to determine how all of these processes would fit together for specific models.

\section{A Theory of ExoDDDM}
\label{sec:model}

ExoDM scenarios have been considered previously in the context of standard halo DM direct detection, and studies have shown that DM candidates with a cosmologically long-lived excited state can arise in complete theoretical constructions \cite{ArkaniHamed:2008qn,Batell:2009vb,Essig:2010ye,Graham:2010ca}.  However, since models of DDDM require a number of fields and involved dynamics we will present a model of ExoDDDM here as proof-of-principle.  To do this we must combine the ingredients of ExoDM models \cite{ArkaniHamed:2008qn,Batell:2009vb,Essig:2010ye,Graham:2010ca} with those of DDDM models \cite{Fan:2013yva,Fan:2013tia}.

We begin in the UV with two Abelian gauge symmetries, $\U(1)' \times \U(1)_D$.  The $\U(1)'$ will be spontaneously broken by a scalar Higgs field $H'$ at the GeV-scale leading to a massive dark gauge boson $Z'$.  We assume this dark gauge boson is kinetically mixed with hypercharge and has mass mixing with the Z-boson \cite{Babu:1997st}, enabling it to mediate exothermic DDDM scattering on nuclei following standard constructions with light mediators \cite{ArkaniHamed:2008qn,Batell:2009vb,Essig:2010ye,Graham:2010ca,Frandsen:2011cg}.\footnote{We thank Felix Kahlhoefer for discussions on this point.}

The $\U(1)_D$ will be unbroken, leaving a massless dark photon $\gamma_D$ which enables the efficient cooling of the DDDM to form a disk structure as in \cite{Fan:2013yva,Fan:2013tia}.  Unlike the $\U(1)'$ we require no kinetic mixing between $\U(1)_D$ and $\U(1)_Y$.  This can be enforced if the $\U(1)_D$ is embedded in a non-Abelian gauge symmetry in the UV \cite{Dienes:1996zr}, or by other means.\footnote{See e.g.\ the discussion in \Refs{Fan:2013yva,Fan:2013tia}.}

\begin{table}[ht]
\caption{ExoDDDM Gauge Charges}
\centering
\begin{tabular}{c | c c c c}
 & $C$&$Y_1$&$Y_2$ & $H'$ \\
\hline
$U(1)'$ &$0$&$1$&$-1$ & $2$ \\
$U(1)_D$ & $1$ & $1$ & $1$ & $0$
\end{tabular}
\label{tab:charges}
\end{table}

The gauge eigenstates in the matter sector are Dirac fermions, $C,Y_1,Y_2$, with charges as detailed in \Tab{tab:charges}.  The Lagrangian is given by
\begin{eqnarray}
\mathcal{L} & = & \mathcal{L}_{\text{SM}} + \epsilon' F'_{\mu \nu} F^{\mu \nu} + \delta m^2 Z'_{\mu} Z^{\mu} + \mathcal{L}_{\text{Kin}} - V \left(H_D,{H_D}^\ast \right) \\
& &  - m_C \overline{C} C - m_Y ( \overline{Y}_1 Y_1 + \overline{Y}_2 Y_2) -  (\lambda H' \overline{Y}_1 Y_2 + \text{h.c.})  ~~,
\label{eq:lag}
\end{eqnarray}
where we have given equal vector-like mass to both $Y$ fermions.\footnote{The equality of masses can enforced by some symmetry in the UV, such as an $\SU(2)$ gauge symmetry of which $Y_1$ and $Y_2$ form a doublet.  This could be broken by an $\SU(2)$ adjoint to $\SU(2) \rightarrow \U(1)'$, leading to the charges of \Tab{tab:charges}.  In this case $H'$ could arise as the off-diagonal component of an $\SU(2)$ adjoint, which breaks $\U(1)' \rightarrow \emptyset$.}  We assume that the scalar potential is minimized with non-zero vacuum expectation for $\langle H' \rangle \sim $ GeV generating a mass for the $Z'$ boson in the $M_{Z'} \sim 100$ MeV range.  We also assume that the Yukawa couplings are small, and after $\U(1)'$ breaking we have $\langle \lambda H' \rangle = \delta/2 \sim 10\text{'s keV}$.\footnote{These small Yukawa couplings of $\lambda \sim \mathcal{O} (10^{-5})$ are technically natural, and so we put them in by hand at this value, as in the SM.  However it would be interesting to consider generating them through some mechanism, originating perhaps at the loop level.}  Including this symmetry breaking the gauge eigenstates $Y_{1,2}$ will mix, forming $\U(1)_D$ eigenstates of mass $m_\pm = m_Y \pm \delta/2$, which we will denote $X_{\pm}$.  After diagonalizing to the mass eigenstate basis the dark photon $\gamma_D$ couples diagonally and the dark $Z'$ boson entirely off-diagonally,
\be
\mathcal{L} \supset \sum_\pm g_D \overline{X}_\pm \cancel{\gamma}_D X_\pm + g' \overline{X}_\pm \cancel{Z}' X_\mp  ~~.
\ee

Thus the matter sector of the DDDM will consist of a light Dirac fermion $C$ which can be thought of as the DDDM ``electron'' \cite{Fan:2013yva,Fan:2013tia}, and the two heavy Dirac fermions $X_{\pm}$ which can be thought of as DDDM ``protons''.  All have unit charge under the $\U(1)_D$ symmetry and are subject to the long-range forces it mediates.

Before we consider cosmology and direct detection it is pertinent to establish the lifetime of these states.  Both $C$ and $X_-$ are absolutely stable due to symmetries analogous to lepton and baryon number, however the situation is more subtle for $X_+$.  The dark photon couples only diagonally to the $X_\pm$ mass eigenstates, and so the decay process $X_+ \rightarrow X_- + n\times \gamma_D$ is forbidden at tree-level.\footnote{Since the $Z'$ couples entirely off-diagonally this process does not arise at one-loop level either.  This is because any decay $X_+ \rightarrow X_- + n\times \gamma_D$ requires an incoming $X_+$ and outgoing $X_-$.  In any loop diagram an internal vertex involving a $Z'$ boson changes $X_+ \rightarrow X_-$, however there are an even number of such vertices since each internal $Z'$ has two endpoints, and no net change leading to $X_+ \rightarrow X_-$ can be generated.  If the $Z'$ coupled both off-diagonally and diagonally this would not be the case and loops could generate $X_+ \rightarrow X_- + n\times \gamma_D$, hence the assumption of equal vector-like masses in \Eq{eq:lag} is critical to the lifetime of the excited state.}  The decay $X_+ \rightarrow Z' + X_- $ is kinematically forbidden, however the decays $X_+ \rightarrow X_- + \overline{\nu} \nu$ and $X_+ \rightarrow X_- + 3 \gamma$ are generated due to mixing between the $Z'$ boson and the SM photon and $Z$-boson.  The lifetime for these decays has been calculated in \cite{Batell:2009vb} for the parameters of interest and comfortably exceeds the age of the Universe.  De-excitation in the early Universe which would deplete the number density of excited states is inefficient \cite{Batell:2009vb} and the current relic abundance of $X_+$ is similar to $X_-$.

The dark photon $\gamma_D$ drives efficient annihilation of the light fermion $C$ in the early Universe, washing out any relic abundance \cite{Fan:2013yva,Fan:2013tia}.  However this state is required for the cooling and eventual collapse into a dark disk, so we follow \cite{Fan:2013yva,Fan:2013tia} and assume that in the early Universe a number asymmetry is generated in $C$ fermions along with an opposite asymmetry in $X_\pm$.  This is analogous to the generation of the baryon asymmetry and, given the plethora of successful models that can generate such an asymmetry in the dark sector  \cite{asymm}, we consider this to be a reasonable assumption.  Since this asymmetry is shared equally between $X_\pm$ we have  $n_{\overline{C}} = 2 n_{X_+} = 2 n_{X_-}$ and there is no net $\U(1)_D$ charge asymmetry.  Thus an asymmetric component of $X_\pm$ will exist in this model, however there may also be an additional symmetric component, depending on the coupling strength of the  $\U(1)_D$ and $\U(1)'$.

Finally we must determine the direct detection cross-section.  The ExoDDDM-nucleon cross-section in the elastic scattering limit is given by
\begin{eqnarray}
\sigma_n & = & 16 \pi \alpha' \alpha_{EM} {\epsilon'}^2 \frac{\mu_n^2}{M_{Z'}^4} \\
& \approx & \left( \frac{\epsilon'}{10^{-6}} \right)^2 \left( \frac{\alpha_D}{10^{-4}} \right) \left( \frac{100 \text{ MeV}}{M_{Z'}} \right)^4 \times 1.4 \times 10^{-40} \text{ cm}^2 ~~,
\end{eqnarray}
hence, due to the light mediator, the required direct detection cross-section is easily obtained even for extremely small kinetic mixing.  This completes the model, which contains all of the necessary ingredients for ExoDDDM and is consistent with current bounds.

\section{Conclusions}
\label{sec:conclusions}
The idea that the entire DM abundance consists of a single species of cold and collisionless particle has dominated thinking in DM research for a long time.  An alternative possibility, where there also exist subdominant components of the total DM abundance which exhibit rich and complex dynamics is relatively unexplored, even though it is both plausible and interesting.  One recently proposed concrete scenario, ``Double-Disk Dark Matter'' \cite{Fan:2013yva,Fan:2013tia} involves a component of DM which has long range interactions and can cool to form complex structures such as galactic DM disks, in analogy with the behavior of visible matter.  DDDM has many novel phenomenological signatures.  However in the simplest scenarios the direct detection prospects appear limited \cite{Fan:2013yva,Fan:2013tia}.  

In this work we have demonstrated that if the DDDM contains excited states which can scatter exothermically on nuclei (ExoDDDM) the direct detection phenomenology can instead be very rich, leading to novel signatures that could distinguish DDDM signals from standard DM candidates.  The signatures particular to ExoDDDM include highly peaked recoil spectra, a reduced annual modulation amplitude and, if a large number of events were accumulated and the modulation amplitude has not been suppressed too greatly due to the exothermic splitting, an unexpected phase of the annual modulation.

As well as outlining the broad qualitative features of ExoDDDM we have also calculated current direct detection limits on ExoDDDM and investigated whether any of the direct detection anomalies could be explained by such a scenario.  Intriguingly the $\sim 3 \sigma$ excess recently announced by the CDMS collaboration \cite{Agnese:2013rvf} can be well explained with ExoDDDM scattering, in some cases with the majority of the preferred parameter space  unconstrained by limits from other experiments.  We have demonstrated that an ExoDDDM interpretation of the CDMS-Si excess is consistent with collider and indirect limits, and have also sketched a simple model which accommodates exothermic scattering on nuclei and sufficiently rapid cooling and collapse of DM into a dark disk.

As always with anomalous events near threshold in direct detection experiments, only time and further experimental investigation, including a push to understand lower nuclear recoil thresholds, will ultimately determine whether the CDMS-Si excess is unexpected background or tentative hints of DM signal.  In the latter case ExoDDDM offers a novel and self-consistent interpretation.

\acknowledgments{We are grateful to Jiji Fan, Patrick Fox, Andrey Katz, Christopher McCabe, Matthew Reece, and Jessie Shelton for useful discussions and to Adam Anderson, Julien Billard, Felix Kahlhoefer and Neal Weiner for useful conversations and comments on an early version of the draft.  M.M. is supported by a Simons Postdoctoral Fellowship.  The work of L.R. was supported in part by by the Fundamental Laws Initiative of the Harvard Center for the Fundamental Laws of Nature and by NSF grants PHY-0855591 and PHY-1216270}

\appendix
\section{Experimental Details}
\label{sec:details}
We briefly review the parameters and data used to produce the experimental limits and best-fit plots of \Sec{sec:quantitative}.  The parameters and methods used are very similar to those found in \cite{Brown:2011dp,Frandsen:2013cna,DelNobile:2013cta}.  We use a Maxwellian velocity distribution for both the SHM and DDDM.  For the SHM we use a rotational velocity, velocity dispersion, and escape velocity of $v_r = 220 \text{ km/s}$, $v_0 = 220 \text{ km/s}$, $v_e = 544 \text{ km/s}$.  For DDDM we use $v_r = 0 \text{ km/s}$, $v_0 = 25 \text{ km/s}$, $v_e = 544 \text{ km/s}$.  We use the standard parameterization of the Solar and Earth velocities \cite{Lewin:1995rx}.  For all target nuclei we use the Helm form factor.

\subsection*{Limits}
\begin{itemize}
\item  XENON10:  The S2-only analysis \cite{Angle:2011th} is used, with the ionization yield $\mathcal{Q}_y$ also taken from \cite{Angle:2011th}.  The energy resolution is assumed to be $\Delta E_R = 1 / \sqrt{2 E_R}$, the acceptance is $95\%$, and the exposure is $15$ kg days.  We use Yellin's ``Pmax'' method \cite{Yellin:2002xd} to set limits.  Although there were previously discrepancies between the limits found in \cite{Angle:2011th} and independent analyses \cite{Farina:2011pw,Cline:2012ei,Frandsen:2013cna}, they have been resolved in an update to \cite{Angle:2011th}.  We find excellent agreement with the new limits in \cite{Angle:2011th}, but slightly stronger limits than \cite{Frandsen:2013cna}.  This is because the ``Maximum Gap'' method \cite{Yellin:2002xd} was used in \cite{Frandsen:2013cna}, giving weaker limits than one finds with the Pmax method used here and in \cite{Angle:2011th}.

\item  XENON100:  We use the results from $20.9$ kg years of running \cite{Aprile:2012nq} with the $\mathcal{L}_{eff}$ measurements from \cite{Aprile:2013teh}.\footnote{It should be noted that lower values of $\mathcal{L}_{eff}$ could bring the XENON100 limits into agreement with the a light elastically scattering DM interpretation of the CDMS-Si excess \cite{Hooper:2013cwa}.}  We convert recoil energy to S1 signal using Eq.\  (1) of \cite{Aprile:2013teh}.  This is then used to calculate the energy resolution based on Poisson fluctuations.   The Maximum Gap method is used to set limits.

\item  CDMS-Ge:  We use the low-threshold study of \cite{Ahmed:2010wy} and only consider limits from the most constraining detector, T1Z5.  Events and efficiencies are available within the supplementary material accompanying \cite{Ahmed:2010wy}.  We use an energy resolution of $\Delta E_R = 0.2 \sqrt{E_R}$ \cite{SchmidtHoberg:2009gn}.  Limits are calculated using the ``Pmax'' method \cite{Yellin:2002xd}.

\item  SIMPLE:  We use the Stage 2 results \cite{Felizardo:2011uw}, of the $C_2 Cl F_5$ detector with $6.71$ kg days exposure.   We take the cut acceptance and nucleation efficiency from \cite{Felizardo:2011uw}.  We use the Feldman-Cousins method \cite{Feldman:1997qc} to determine a $90\%$ upper limit of $2.38$ signal events based on an expected background of $2.2$ events and the observation of one event above the $8$ keV threshold.

\item  CRESST-II Commissioning Run:  We use the analysis of \cite{Brown:2011dp} which includes possible scattering on oxygen and calcium.  Exposures, acceptances, and resolutions for the Zora/SOS23 and Verena/SOS21 detectors can be found in \cite{Brown:2011dp}.  We use the Maximum Gap method \cite{Yellin:2002xd} to set limits.

\end{itemize}

\subsection*{Signal Hints}

\begin{itemize}
\item  CDMS-Si: We use the three events found in presented in $140.2$ kg days of data \cite{Agnese:2013rvf}.  We assume a detector resolution of 0.3 keV \cite{Akerib:2010pv} and acceptance from \cite{Agnese:2013rvf}.  We take the background contributions from \cite{APStalk} and choose a normalization such that surface events, neutrons, and ${}^{206}$Pb, give  $0.41$, $0.13$, and $0.08$ events respectively.  We employ the extended maximum likelihood method \cite{Barlow:1990vc} to determine best-fit parameters.

\item  CRESST-II:  We use the results of \cite{Angloher:2011uu} with events in bins of $3$ keV width as in \cite{Frandsen:2011gi}.  The acceptance is assumed to be $A_O (86\%)$, $A_{Ca} (90\%)$,  and $A_W (89\%)$ and total exposure is $730$ kg days.  In each bin we calculate a Poisson probability for the observed number of events as a function of total signal and estimated background events, and then construct a likelihood to estimate best-fit parameters.

\item  DAMA:  We use the combined DAMA/NaI and DAMA/LIBRA event rates \cite{Bernabei:2010mq} and only use the lowest $8$ bins as the higher-energy bins do not exhibit significant modulation.  We use a quenching factor of $q_{Na} = 0.3$.

\item  CoGeNT:  We only consider the modulated data, since this constitutes the strongest hint for a DM signal.  We use the publicly available data from \cite{Aalseth:2011wp} and the detector resolution and quenching factor from \cite{Aalseth:2008rx} and consider modulation data binned as in \cite{Frandsen:2013cna}.
\end{itemize}

\bibliographystyle{JHEP}

\end{document}